\documentclass[twocolumn,showpacs,preprintnumbers,amsmath,amssymb,superscriptaddress,prx,longbibliography]{revtex4-1}

\usepackage{graphics}
\usepackage{subfigure}
\usepackage[normalem]{ulem}
\usepackage{hyperref}
\hypersetup{colorlinks=true, citecolor=blue, urlcolor=blue, linkcolor=blue}

\allowdisplaybreaks

\begin{document}

\title{Deep Inelastic Scattering on Ultracold Gases}

\author{Johannes Hofmann}
\affiliation{TCM Group, Cavendish Laboratory, University of Cambridge, Cambridge CB3 0HE, United Kingdom}

\author{Wilhelm Zwerger}
\affiliation{Technische Universit\"at M\"unchen, Physik Department, James-Franck-Strasse, 85748 Garching, Germany and
Theoretische Physik, ETH Z\"urich, 8093 Z\"urich, Switzerland}

\date{\today}

\begin{abstract}
We discuss Bragg scattering on both Bose and Fermi gases with strong short-range interactions in the deep inelastic regime of large wave vector transfer $q$, where the dynamic structure factor is dominated by a resonance near the free-particle energy $\hbar\omega=\varepsilon_{\bf q}=\hbar^2 q^2/2m$.  Using a systematic short-distance expansion, the structure factor at high momentum is shown to exhibit a nontrivial dependence on frequency characterized by two separate scaling regimes. First, for frequencies that differ from the single-particle energy by terms of order ${\cal O}(q)$ (i.e., small deviations compared to the single-particle energy), the dynamic structure factor is described by the impulse approximation of Hohenberg and Platzman. Second, deviations of order ${\cal O}(q^2)$ (i.e., of the same order or larger than the single-particle energy) are described by the operator product expansion, with a universal crossover connecting both regimes. The scaling is consistent with the leading asymptotics for a number of sum rules in the large momentum limit. Furthermore, we derive an exact expression for the shift and width of the single-particle peak at large momentum due to interactions, thus extending a result by Beliaev [J. Exp. Theor. Phys. {\bf 7}, 299 (1958)] for the low-density Bose gas to arbitrary values of the scattering length~$a$. The shift exhibits a maximum around $qa \simeq 1$, which is connected with a maximum in the static structure factor due to strong short-range correlations. For Bose gases with moderate interaction strengths, the theoretically predicted shift is consistent with the value observed by Papp {\it et al.} [Phys. Rev. Lett. {\bf 101}, 135301 (2008)]. Finally, we develop a diagrammatic theory for the dynamic structure factor which accounts for the correlations beyond Bogoliubov theory. It covers the full range of momenta and frequencies and provides an explicit example for the emergence of asymptotic scaling at large momentum.
\end{abstract}

\maketitle

% 05.30.Jp	Boson systems
% 03.75.Hh	Static properties of condensates; thermodynamical, statistical, and structural properties
% 67.85.Bc	Static properties of condensates
% 67.85.De	Dynamic properties of condensates; excitations, and superfluid flow
% 67.85.Jk	Other Bose-Einstein condensation phenomena
% 31.15.-p	Calculations and mathematical techniques in atomic and molecular physics
% 34.50.-s	Scattering of atoms and molecules

\section{Introduction}

Deep inelastic scattering --- the inelastic scattering off a target with a high-energy probe --- is a central theme in high-energy physics. It has played a crucial role in establishing quarks as the basic fields for describing the internal structure of hadrons~\cite{taylor91,kendall91,friedman91}. Originally, the theoretical analysis of deep inelastic scattering was based on the parton model due to Bjorken and Paschos~\cite{bjorken69} and Feynman~\cite{feynman69}, in which a virtual photon created (e.g., in electron scattering) sees the constituents (partons) inside the hadron as quasifree particles because the time scale $1/cq$ of the virtual photon interaction is much shorter than the parton interaction time. The measured structure functions are then proportional to the density of partons with a certain fraction of the nucleon momentum~\cite{peskin95}. In particular, dimensionless ratios turn out to be asymptotically scale invariant, depending only on the Bjorken variable $X=\hbar q^2/2m\omega$, where $\hbar q$ and $\hbar \omega$ denote the momentum and energy transfer by the photon, and $m$ is the hadron mass~\cite{peskin95}. In contrast to high-energy physics, research in condensed matter and many-body physics is traditionally concerned with low-energy, long-distance phenomena. In the case of dilute, ultracold quantum gases, these may be described by simple models such as Bogoliubov's weakly interacting Bose fluid, which --- at vanishing chemical potential --- defines a scale-invariant field theory, giving rise to universal behavior at low densities and temperature~\cite{*[{}] [{, Chap. 16.}] sachdev11}. Short-distance physics, in turn, depends on the details of the short-range interaction and is therefore not expected to exhibit universal behavior.

In our present work, we show that scaling akin to that studied in deep inelastic scattering in high-energy physics also appears in ultracold gases. Specifically, we study the dynamic structure factor in the regime of high momenta. In this regime, the assumption that interactions between the atoms are negligible leads to the so-called impulse approximation (IA)~\cite{hohenberg66}, which may be viewed as an analog of the parton model in high-energy physics. As we show, this approximation corresponds to a quasifree regime that is dominated by single-particle excitations. It leads to a particular form of scaling, yet fails to account for multiparticle excitations. The latter can be incorporated in a systematic manner using the Wilson operator product expansion (OPE). The simple scaling in the quasifree regime is then replaced by a more complicated one involving anomalous dimensions, reminiscent to what is achieved in high-energy physics with the QCD-improved parton model~\cite{altarelli08}.

In the context of strongly interacting quantum fluids, much of our understanding of their excitation spectrum is derived from the dynamic structure factor, which determines the scattering rate of an external density probe that transfers an energy $\hbar \omega$ and momentum $\hbar {\bf q}$ to the system. It is measured by inelastic neutron scattering in $^4$He~\cite{woods73,griffin93,snow95,griffin95,pitaevskii16} or via two-photon Bragg spectroscopy in ultracold quantum gases~\cite{kozuma99,stenger99,stamperkurn99,steinhauer02,papp08,veeravalli08}. The dynamic structure factor $S(\omega, {\bf q})$ is defined  through the imaginary part of the density response function $\chi$~\footnote{Note that compared to Ref.~\cite{pitaevskii16} we include an additional factor of volume. Our structure factor thus has dimension of inverse energy and volume.}:
\begin{align}
S(\omega, {\bf q}) &= \frac{1}{\pi} \bigl[1-e^{-\beta \hbar \omega}\bigr]^{-1} \, {\rm Im} \, \chi(\omega + i0, {\bf q}) .
\end{align}
At small momentum transfer, $S(\omega, {\bf q})$ is dominated by collective excitations. As shown by Feynman~\cite{feynman54}, these are phonons with a linear dispersion $\omega_{\bf q}=c_sq$, where $c_s$ is the speed of sound. In the superfluid phase, phonons in fact exhaust the $f$-sum rule,
\begin{align}
m_1 &= \hbar^2 \int_{-\infty}^\infty \, d\omega \, \omega S(\omega, {\bf q}) = n \varepsilon_{\bf q} , \label{eq:fsumrule}
\end{align}
in the long-wavelength limit $q\to 0$ (here, $\varepsilon_{\bf q}=\hbar^2q^2/2m$ is the free particle energy and $m$ the bare mass of an atom). As a result, the density response function has a single pole at a position $\hbar \omega_{\bf q} = \varepsilon_{\bf q}/S({\bf q})$, which is fixed by the static structure factor $S({\bf q})$ via the sum rule $m_0= n S({\bf q})$. This is the so-called single-mode approximation, $S(\omega, {\bf q}) \to S_{1p}(\omega, {\bf q}) = Z_{\bf q} \delta(\hbar \omega - \hbar \omega_{\bf q})$, which is exact at low momenta. In this limit, both the excitation frequency $\omega_{\bf q}$ and the quasiparticle weight $ Z_{\bf q}=n S({\bf q})$ depend only on the single parameter $c_s$, which is fixed by the compressibility. As a consequence, the single-mode approximation does not provide any information about superfluid properties such as the superfluid or the condensate density~\footnote{In fact, as shown by Wagner~\cite{wagner66}, the presence of phonon-like excitations in the long wavelength limit of $S(\omega, {\bf q})$ is insensitive to the existence of a broken gauge symmetry, which requires additionally that the phonons also appear as sharp poles of the {\it single}-particle Green's function.}. It is therefore of considerable interest to study which kind of information is contained in the dynamic structure factor away from the long-wavelength limit. Now, as argued by Feynman~\cite{feynman54}, a simple extension of the  single-mode approximation to larger wave vectors leads, in the particular case of $^4$He, to a roton minimum in the excitation energy $\hbar \omega_{\bf q}$. This is a consequence of the pronounced peak in the static structure factor $S({\bf q})$ near the wave vector $q_0\simeq 2\, \text{\normalfont\AA}^{-1}$ associated with the short-range order in the strongly correlated fluid. Quantitatively, however, the single-mode approximation estimate for  the excitation energy $\hbar \omega_{{\bf q}_0}$ near $q_0$ is a factor of $2$ larger than the experimental result~\cite{pitaevskii16}.  The physics behind the breakdown of the single-mode approximation has been discussed by Miller, Pines, and Nozi\`{e}res~\cite{miller62}: they have shown that the backflow corrections to the Feynman variational ansatz $|\psi_{\bf q}\rangle=\hat{\rho}_{\bf q}^{\dagger}|0\rangle$ for excited states with wave vector ${\bf q}$ as well as the strong depletion of the condensate become increasingly important at larger wave vectors, giving rise to an incoherent background $S_{\rm inc}(\omega, {\bf q})$. Its integrated weight $m_0^{\rm inc}=n S({\bf q})[1-f({\bf q})]$ defines a dimensionless function $f({\bf q})$ that approaches unity as $f({\bf q})\to 1- \mathcal{O}(q^4)$ in the long-wavelength limit but vanishes quickly beyond wave vectors of the order of the inverse interparticle spacing.  In this regime, the dynamic structure factor is dominated by an incoherent background which depends on microscopic details. Surprisingly, however,  in the regime of very large momenta $q\gg q_0$, a completely different kind of universality emerges.  Indeed, as anticipated by Miller, Pines, and Nozi\`{e}res~\cite{miller62} and then shown in detail by Hohenberg and Platzman~\cite{hohenberg66}, the dynamic structure factor at large wave vectors provides a direct measure of the momentum distribution. It thus allows to infer the presence of a nonvanishing condensate density $n_0$ and the associated off-diagonal long-range order in an interacting Bose fluid. This prediction is based on the so-called impulse approximation, which assumes that at large wave vectors ${\bf q}$, the response is given by a Fermi golden rule expression for exciting a {\it single} atom with small momentum ${\bf k}$ to a large momentum ${\bf k} + {\bf q}$. Neglecting interactions between the final and initial state atoms, this yields the IA
\begin{align}
S_{\rm IA}(\omega, {\bf q}) &= \int \frac{d^3k}{(2\pi)^3} \, n({\bf k}) \, \delta(\hbar \omega + \varepsilon_{\bf k} - \varepsilon_{\bf k+q}) , \label{eq:defIA}
\end{align}
in which the dynamic structure factor is completely determined by the momentum distribution $n({\bf k})$ of the strongly interacting quantum fluid. This approximation may be viewed as analogous to the naive parton model of high-energy physics where the structure functions are proportional to the density of different partons that carry a certain fraction of the nucleon momentum~\cite{bjorken69,peskin95}. A crucial prediction of the IA is a particular form of scaling: $S(\omega, {\bf q})$ does not depend on $\hbar \omega$ and ${\bf q}$ separately but only on a single dimensionless scaling variable. Specifically, assuming  a rotationally invariant system with a finite condensate density $n_0$, the general form 
\begin{align}
n({\bf k}) = (2\pi)^3 n_0 \delta({\bf k}) + \tilde{n}(k)  \label{eq:condensate}
\end{align}
of  the momentum distribution implies
\begin{align}
S(\omega, {\bf q}) &= \frac{m}{\hbar^2 \tilde{\xi}^2} \frac{1}{q} J_{\rm IA}(Y) , \quad {\rm with}\, 
\quad Y = \frac{m \tilde{\xi}}{\hbar^2} \frac{\hbar \omega - \varepsilon_{\bf q}}{q} ,\label{eq:IA}
\end{align}
where $Y$ is sometimes referred to as the West scaling variable~\cite{west75,griffin93}~\footnote{Note that our definition of $Y$ differs from the literature~\cite{snow95,pitaevskii16}, where the impulse approximation scaling variable is often defined as $Y_{\rm lit} = m (\hbar \omega - \varepsilon_{\bf q})/\hbar^2 q=Y/\tilde{\xi}$, i.e., it has dimension of an inverse length.}. The associated scaling function  
\begin{align}
J_{\rm IA}(Y) &= n_0 \tilde{\xi}^3 \delta(Y) + \frac{\tilde{\xi}^2}{4 \pi^2} \int_{|Y|/\tilde{\xi}}^\infty dk \, k \, \tilde{n}(k) \label{eq:JIA}
\end{align}
contains a singular contribution in the presence of a finite condensate plus a smooth part, which reflects the momentum distribution $\tilde{n}(k)$ of noncondensed atoms. The dimensionless scaling variable $Y$ in Eq.~\eqref{eq:IA} involves a length scale $\tilde{\xi}$ whose inverse is the characteristic scale over which the momentum distribution varies.  For weakly interacting bosons, a convenient choice for $\tilde{\xi}$ is thus the standard healing length $\xi$, which appears in Bogoliubov theory. For both degenerate Fermi gases or for strongly interacting bosons, in turn, the momentum distribution has the inverse $1/\tilde{\xi}\simeq n^{1/3}$ of the average interparticle spacing as a characteristic wave number scale, while for nondegenerate gases a convenient choice for $\tilde{\xi}$ is  the thermal wavelength $\lambda_T$. To be consistent with the $f$-sum rule~[Eq.~\eqref{eq:fsumrule}], the smooth part of the scaling function $J_{\rm IA}(Y)$ away from the condensate peak, which is called the Compton profile or the longitudinal momentum distribution in the $^4$He literature~\cite{snow95}, must take up the missing area $n-n_0$. Because of a strong condensate depletion, this is quite large in $^4$He -- close to 90\% even at zero temperature. Neutron scattering in the regime of large momentum transfer $q$ provides quantitative results for the smooth part of the scaling function $J_{\rm IA}(Y)$~\cite{snow95}.  Because of the finite instrumental resolution and the unknown final-state effects which -- as we show below -- limit the range of applicability of the IA to $|Y|\ll \mathcal{O}(q^1)$,  the extracted values for the condensate density of $^4$He have considerable error bars. They are consistent, however, with the accepted theoretical result $n_0(T=0)\simeq 0.1\, n$, which relies on path-integral or Green's function Monte Carlo simulations based on ab initio pair potentials~\cite{ceperly95,giorgini92}. 

The realization of a completely novel class of Bose-Einstein condensates (BEC) using ultracold alkali gases~\cite{davis95,anderson95}  has opened new opportunities to study both collective and single-particle excitations of superfluids~\cite{pitaevskii16}.  In the ultracold limit, the interactions in these gaseous systems are completely specified by the $s$-wave scattering length $a$. For bosons  in three dimensions, stability requires $a$ to be positive, whereas both signs of $a$ are possible for two-component Fermi gases~\cite{zwerger16}. In the absence of a Feshbach resonance, the characteristic values of the scattering length are of the order of the van der Waals length $\ell_{\rm vdW}$, which is typically in the few nanometer range.  Both the average interparticle spacing $n^{-1/3}$ and the wavelengths $4\pi/q$ used in Bragg spectroscopy then obey $n^{-1/3}\gg |a|$ and $1/q \gg |a|$. In this regime of weak correlations, Bose gases  are well described by the standard Bogoliubov theory, which is based on the assumption of a classical coherent state that represents the condensate.  The Gaussian fluctuations on top of the condensate then give rise to a set of noninteracting quasiparticles. Their spectrum $E_{\bf q}=\sqrt{\varepsilon_{\bf q}(\varepsilon_{\bf q}+2gn_0)}$ is linear in momentum $E_{\bf q}\to \hbar c_sq$ below the inverse healing length $1/\xi$ and approaches the free-particle limit as $E_{\bf q}=\varepsilon_{\bf q}+gn+\cdots$ at large wave vectors $q\xi\gg 1$. Here, $g=4\pi\hbar^2 a/m$ is  the low-energy coupling constant, linear in the scattering length $a$. Within Bogoliubov theory, the bosonic quasiparticles with spectrum $E_{\bf q}$ exhaust the complete set of excitations. As a result, the single-mode approximation turns out to be exact at arbitrary wave vectors $q$. Such a simple description, however, is no longer applicable once the scattering length is increased up to values of the order of or even larger than either  $n^{-1/3}$ or $1/q$. This is possible via Feshbach resonances~\cite{chin10}. The use of Feshbach resonances to study strongly interacting gases has been particularly successful for two-component Fermi gases, which are stable with respect  to three-body losses near the unitary limit of infinite scattering length~\cite{zwerger16}. Bose gases, unfortunately, do not enjoy this stability since the decay rate due to three-body losses increases like $\Gamma_3\sim \hbar n^2a^4/m$ on average~\cite{fedichev96,nielsen99}. More precisely, for large scattering lengths, Bose gases are unstable due to the presence of the Efimov effect, i.e., the formation of three-body bound states at both positive and negative scattering lengths. For open channel dominated Feshbach resonances, this happens in a regime $|a|\gtrsim 10\,\ell_{\rm vdW}$~\cite{wang12,schmidt12}.
 
Nevertheless, a number of experiments in recent years have explored Bose gases with scattering lengths larger than the average interparticle spacing or the inverse thermal wavelength $\lambda_T$~\cite{rem13,fletcher13,makotyn14}. Regarding the dynamic structure factor, the failure of Bogoliubov theory in the regime $q|a|=\mathcal{O}(1)$ was observed some time ago in a Bragg scattering experiment on $^{85}$Rb by Papp {\it et al.}~\cite{papp08}. The experiment measures the so-called line shift $\Delta (\hbar \omega) = \hbar \omega_{\bf q} - \varepsilon_{\bf q}$, which is the deviation of the peak position at $\hbar \omega_{\bf q}$ in the dynamic structure factor from the single-particle energy $\varepsilon_{\bf q}$. Within Bogoliubov theory, the line shift is given by the mean-field energy $\Delta (\hbar \omega) = g n$ of the gas. It is linear in both the scattering length and the total number density $n$ since the depletion $n-n_0 \sim \sqrt{na^3}$ of the condensate by interactions is of higher order in the small parameter $na^3\ll 1$. The measurement~\cite{papp08} is carried out at fixed large momentum as a function of the scattering length, and, indeed, the linear-in-$a$ Bogoliubov behavior is found experimentally for $qa\ll 1$. With increasing scattering length, however, the observed shift reaches a maximum for values $qa=\mathcal{O}(1)$ and then starts to decrease.

%++++++++++++++++++++++++++++++++++++++++
\begin{figure}[t!]
\scalebox{0.6}{\includegraphics{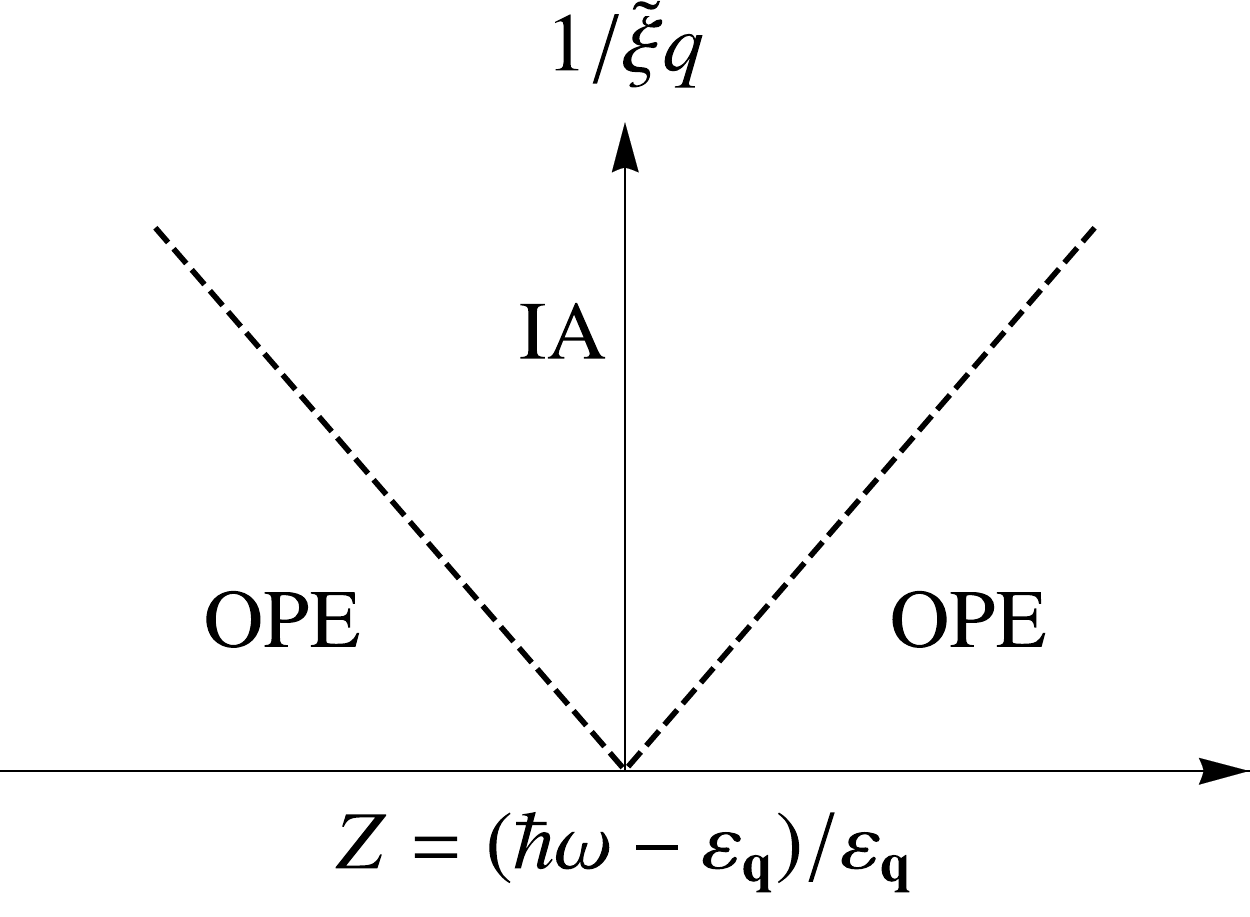}}
\caption{Sketch of the asymptotic structure of the dynamic structure factor at large momentum $q\tilde{\xi} \gg 1$ and $q|a| \gg 1$, where $\tilde{\xi}$ is the characteristic length scale of the gas [such as $k_n^{-1}=(6\pi^2n)^{-1/3}$ or $\lambda_T = \hbar/\sqrt{2\pi m T}$]. Note that this scaling does not necessarily require $k_n |a| \gg 1$. Small deviations in energy from the single-particle peak of order $\mathcal{O}(q)$ are described by the impulse approximation (IA), whose range of applicability shrinks with increasing momentum. Large-energy deviations of order $\mathcal{O}(q^2)$ are in turn described by the operator product expansion (OPE), which predicts asymmetric tails on the left- and right-hand side of the single-particle peak.
}
\label{fig:scaling}
\end{figure}
%++++++++++++++++++++++++++++++++++++++++

In our present work, we discuss the dynamic structure factor of both Bose and Fermi gases with strong interactions, focusing, in particular on the so-called  deep inelastic regime of large momentum transfer. As the main result of our work, which is sketched in Fig.~\ref{fig:scaling}, we establish two distinct scaling regions with separate and complementary regions of validity. For frequencies close to the dominant single-particle peak (we make the notion of ``close'' more precise shortly), the dynamic structure factor is described by the IA~[Eq.~\eqref{eq:defIA}] with a scaling as given in Eq.~\eqref{eq:IA}. It involves a delta peak right at the free-particle energy $\varepsilon_{\bf q}$ in the presence of a condensate and a smooth, symmetric background, cf. Eq.~\eqref{eq:JIA}.  In particular, interaction corrections to the naive Fermi golden rule expression~[Eq.~\eqref{eq:defIA}] turn out to vanish in the limit $qa\gg 1$. Away from the single-particle peak, the dynamic structure factor is described by the operator product expansion, which predicts a scaling of the form
\begin{align}
S(\omega, {\bf q}) &= \frac{m \mathcal{C}_2}{\hbar^2 q^3} J_{\rm OPE}(Z) , \quad {\rm with} \, \quad Z = \frac{\hbar \omega - \varepsilon_{\bf q}}{\varepsilon_{\bf q}} .\label{eq:OPEscaling}
\end{align}
It involves the quite different scaling variable $Z$ which is connected to the Bjorken variable $X=\hbar q^2/2m\omega$ of high-energy physics by $Z=1/X -1$. The prefactor $\mathcal{C}_2$ in Eq.~\eqref{eq:OPEscaling} is the Tan two-body contact density~\cite{tan08a,tan08b,tan08c,braaten08}, which is a measure of the probability for two atoms to be at the same point in space. In particular, we establish that both IA and OPE are complementary: the IA describes the dynamic structure factor in a frequency range close to the single-particle peak, with deviations $\hbar \omega - \varepsilon_{\bf q} = \mathcal{O}(q)$ that scale linearly with momentum. This is a simple consequence of the fact that the single-particle excitations described by Eq.~\eqref{eq:defIA} can extend only up to a range of order $q$ beyond the single-particle peak. Formally, the associated scaling variable thus obeys $Y = \mathcal{O}(q^0)$ in its domain of validity. The IA breaks down for frequencies that deviate from the single-particle peak by terms of order $q^2$; i.e., $\hbar\omega-\epsilon_{\bf q}=\mathcal{O}(q^2)$. Such large deviations may appear due to two-particle excitations in which a large momentum $q$ is transferred to two particles. In this multiparticle regime, $Y=\mathcal{O}(q)$ increases linearly with $q$ and, correspondingly, the scaling variable $Z=2Y/q\tilde{\xi}$ becomes of order $1$. As we show below, an exact description of the Bose gas in this multiparticle regime is provided by the OPE~\eqref{eq:OPEscaling}. This situation is illustrated schematically in Fig.~\ref{fig:scaling}.  Most importantly, the two different scaling regimes turn out to be connected in a continuous manner.  As an application of our results, we show that the shift and width of the single-particle peak can be determined asymptotically using the OPE. This provides a straightforward explanation of the experimental results found by Papp {\it et al.}~\cite{papp08}. Our results on the OPE side extend previous work by Son and Thompson~\cite{son10}, Goldberger and Rothstein~\cite{goldberger12}, Nishida~\cite{nishida12}, and one of the present authors~\cite{hofmann11}, as well as those obtained by Wong in an important early paper~\cite{wong77}. \\

In detail, this paper is structured as follows: Section~\ref{sec:asymptotics} discusses the high-momentum limit of the dynamic structure factor and the scaling predictions of both OPE and IA and establishes the main result of our work discussed above. We check our results in Sec.~\ref{sec:sumrules} by computing the first four moment sum rules as well as so-called Borel sum rules with exponential weight factor, which all agree with the exact results derived from the OPE of the density response function. Section~\ref{sec:lineshift} then extends our OPE results to derive an exact expression for the line shift at high momentum, which is proportional to Tan's two-body contact parameter $\mathcal{C}_2$. The line shift has a nonmonotonous dependence on scattering length, in qualitative agreement with the experimental results of Ref.~\cite{papp08}. In the limit of weak interactions, where the perturbative expression for $\mathcal{C}_2$ can be used, our result agrees with a calculation by Beliaev~\cite{beliaev58b} for a weakly interacting Bose gas. Section~\ref{sec:diagrammatics} develops a diagrammatic approximation to the dynamic structure factor that is consistent with various constraints and with the OPE results. We discuss the calculation based on the many-body $T$ matrix in Sec.~\ref{sec:response} and present the results in Sec.~\ref{sec:results}, paying particular attention to the crossover from the low-momentum to the high-momentum regime, where the dynamic structure factor is described by the combined scaling form of IA and OPE. These results provide quantitative predictions for experiments.  The results derived here are valid not only in the condensed phase but at finite temperature as well. As an example, in Sec.~\ref{sec:Ythermal}, we use a recent computation of the momentum distribution of the nondegenerate Bose gas~\cite{barth15} to compute the universal IA scaling form. Section~\ref{sec:fermi} provides the extension of our Bose gas results to Fermi gases. We end with a summary and conclusions in Sec.~\ref{sec:summary}.

\section{High-momentum behavior of the dynamic structure factor}\label{sec:asymptotics}

\begin{figure}[t]
\scalebox{0.8}{\includegraphics{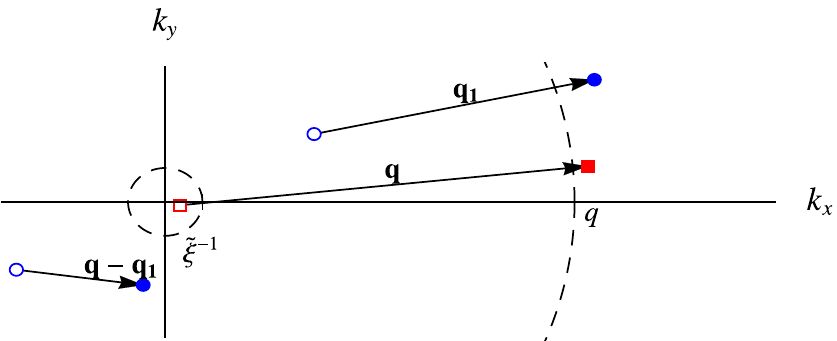}}
\caption[]{High-momentum excitations that determine the deep inelastic form of the dynamic structure factor (in a 2D plane) at large momentum transfer $q \gg \tilde{\xi}^{-1}$, where $\tilde{\xi}$ is the characteristic length scale which sets the typical wave vector of the atoms. We consider two cases: first, a single atom with wave number ${\bf k}$ (empty red square) can be transferred to a state with large wave vector ${\bf k} + {\bf q}$ (red filled square). Second, two initial atoms $({\bf k}',-{\bf k}')$ with large and opposite momenta (blue circles) can be excited to high momenta $({\bf k}' + {\bf q}_1,-{\bf k}' + {\bf q} - {\bf q}_1)$ (blue dots). Energy conservation implies for the probe energy in the first case $\hbar \omega - \varepsilon_{\bf q} = \mathcal{O}(q^1)$ --- the {\it quasifree} regime --- and in the second case $\hbar \omega - \varepsilon_{\bf q} = \mathcal{O}(q^2)$ --- the {\it multiparticle} regime.} 
\label{fig:regimes}
\end{figure}

Both the impulse approximation and the operator product expansion address the short-distance behavior of the density response function
\begin{align}
\chi(\omega, {\bf q}) &= \frac{i}{\hbar V} \int dt \, e^{i \omega t} \langle T_t \hat{n}_{\bf q}(t) \hat{n}_{\bf -q} \rangle \nonumber \\
&= \frac{i}{\hbar} \int dt \int d{\bf r} \, e^{i \omega t - i {\bf q} \cdot {\bf r}} \langle T_t \hat{n}(t,{\bf r}) \hat{n}(0,{\bf 0}) \rangle , \label{eq:defchi}
\end{align}
where $\hat{n}(t,{\bf r})$ is the density operator and $\hat{n}_{\bf q}(t)$ its Fourier transform, and the time evolution of the operators is dictated by the Heisenberg equation of motion, $\hat{n}_{\bf q}(t) = e^{iHt/\hbar} \hat{n}_{\bf q} e^{-iHt/\hbar}$. The difference in both approximations lies in the type of excitation --- single particle for the IA or high-momentum pair excitations for the OPE --- that is taken into account. We sketch this situation in Fig.~\ref{fig:regimes}. A single-particle excitation is created by transferring the large probe wave vector ${\bf q}$ to an initial atom with wave vector ${\bf k}$ (empty red square), which is drawn from an initial distribution $n({\bf k})$ that is concentrated in a momentum range $\tilde{\xi}^{-1} \ll q$. The IA assumes that the high-momentum state (red filled square) propagates as a free particle without interactions. Hence, this regime is called the {\it quasifree} regime. Energy conservation implies $\hbar \omega + \varepsilon_{\bf k} = \varepsilon_{\bf k+q}$, and, hence, the deviation $\hbar \omega - \varepsilon_{\bf q} = \mathcal{O}(q^1)$ of the excitation energy from the single-particle energy scales linearly with wave vector $q$. In addition to these single-particle excitations, it is necessary, however, to consider pair and higher-order excitations in which a large momentum is transferred to two or more particles. The probability for such multiparticle excitations is determined by the likelihood of two or more atoms being close. For just two particles, this may be quantified by the short-distance behavior~\cite{braaten08,zwerger16}
\begin{align}
\lim_{r \to 0} n^2 g^{(2)}(r) = \frac{\mathcal{C}_2}{16\pi^2} \biggl(\frac{1}{r^2}-\frac{2}{ar}+\dots \biggr)  \label{eq:g2}
\end{align}
of the two-particle distribution function, which is simply proportional to the square of the two-body wave function $\psi_0(r)\sim 1/r-1/a$ at zero energy for gases whose interactions are described by a Bethe-Peierls boundary condition. Formally, Eq.~\eqref{eq:g2} follows from the contribution of the contact operator to the operator product expansion in Eq.~\eqref{eq:defOPE} for the special case of equal times $t=0$. For small scattering lengths, the contact density $\mathcal{C}_2(a) =(4\pi na)^2$ vanishes quadratically. In an expansion in powers of $a$, the contribution $-2a/r$ to $g^{(2)}(r)$ is therefore dominant, which is the only one kept within Bogoliubov theory. For $a>0$, this contribution describes the suppression due to repulsive interactions of the probability density to find two particles separated by a distance $r$ smaller than the healing length $\xi$. For separations $r$ smaller than the scattering length, however, the $-2a/r$ contribution is eventually dominated by the term $\mathcal{C}_2/(4\pi nr)^2$ which guarantees that $g^{(2)}(r)$ remains positive at short distances~\footnote{Note that the standard measure $g^{(2)}(0)$ which describes, e.g., the bunching of {\it noninteracting} bosons  in their noncondensed phase via $g_0^{(2)}(0) = 2$  does not exist for gases in the presence of zero-range interactions, where it is essentially replaced by the contact density $\mathcal{C}_2$ as defined in Eq.~\eqref{eq:g2}.}. Provided that the scattering length is much larger than the effective range $\ell_{\rm vdW}$ of interactions, this implies an effective bunching of atoms in a wide range of separations $\ell_{\rm vdW}\ll r<a$. The singular contribution $\sim \mathcal{C}_2/r^2$ to the pair-distribution function was noted first by Naraschewski and Glauber~\cite{naraschweski99} for a weakly interacting Bose gas and was later discussed by Holzmann and Castin~\cite{holzmann99}. While it is difficult to observe for weakly interacting bosons with scattering lengths of order $\ell_{\rm vdw}$, the result~[Eq.~\eqref{eq:g2}] is at least consistent with precision experiments of the -- even time-dependent -- pair-distribution function~\cite{guarrera11}.

%++++++++++++++++++++++++++++++++++++++++
%++++++++++++++++++++++++++++++++++++++++
  
For strongly interacting gases, where the contribution~$\sim~\mathcal{C}_2/r^2$  to the pair-distribution function becomes important, it is necessary to consider the deep inelastic scattering off pairs of atoms with high momenta. Here, interactions must be taken into account, which may redistribute the transferred wave vector ${\bf q}$ between the pair. Such a process is sketched in Fig.~\ref{fig:regimes} by the blue circles (initial state) and blue dots (final state). We call this regime the {\it multiparticle} regime. Here, $\hbar \omega - \varepsilon_{\bf q} = \mathcal{O}(q^2)$, which implies $Y = \mathcal{O}(q^1)$ or $Z=\mathcal{O}(q^0)$ for the scaling variables. Depending on the probe energy $\hbar \omega$, either one of the two types of excitations will dominate the dynamic structure factor. In the following, we discuss the derivation of both IA and OPE and show that they apply in the quasifree and the multiparticle regime, respectively.  Remarkably, there is a smooth crossover that connects both regimes. Taken together, the IA and the OPE therefore provide a complete description of the dynamic structure factor at high momentum, as indicated in Fig.~\ref{fig:scaling}.

To obtain the IA from Eq.~\eqref{eq:defchi} in the quasifree regime, we assume that the time evolution of the density operator $\hat{n}_{\bf q}(t)$ or $\hat{n}(t,{\bf r})$ is governed by the noninteracting Hamiltonian, i.e., $\hat{n}_{\bf q}(t) = \sum_{\bf k} e^{i (\varepsilon_{\bf k-q} - \varepsilon_{\bf k}) t/\hbar} a_{\bf k - q}^\dagger a_{\bf k}$, where $a_{\bf k}^\dagger$ is a Bose creation operator. Similar to the arguments leading to the parton model, this assumption is justified since during the short time scales of the probe, a scattered high-momentum atom is not able to interact with its surroundings. More precisely, the effective collision time $\tau_{\rm sc} = 1/(n \sigma_q v_q)$ must be large compared to the characteristic time scale $\tau_{n} = m/\hbar k_n^2$ set by the finite particle density $n=k_n^3/6\pi^2$ of the Bose gas. Now, in spite of the large velocity $v_q = \hbar q/m$, this assumption holds provided the scattering cross section $\sigma_q$ vanishes faster than $1/q$. In the special case of quantum gases with zero-range interactions, we have (for $q\gg k_n$) $\sigma_q \sim 1/(a^{-2} + q^2)$. For $q|a|\gg 1$, the scattering time is, thus, indeed large, and the IA applies. Note that at high momentum, the condition $q|a|\gg 1$ is weaker than $k_n|a| \gg 1$, with the latter implying the former but not vice versa. If we assume that the probe scatters off an initial atom with small momentum, the product $a_{\bf k - q}^\dagger a_{\bf k'+q}^{}$ of creation and annihilation operators for high-momentum atoms in Eq.~\eqref{eq:defchi} can be replaced by the $c$ number $\delta_{{\bf k},{\bf k}'}$. The remaining expectation value then reduces to the momentum distribution. Performing the time integral  and taking the imaginary part, we obtain the impulse approximation Eq.~\eqref{eq:defIA} with the scaling function $J_{\rm IA}(Y)$ given in Eqs.~\eqref{eq:IA} and~\eqref{eq:JIA}.

Quite generally, the smooth part of the IA scaling function $J_{\rm IA}(Y)$  depends on the details of the momentum distribution, and, hence, on the microscopic low-energy scale. Remarkably, exact results may be derived in the limits $|Y|\ll1$ or $|Y|\gg1$, which hold for arbitrary superfluids or for ultracold gases, respectively. Discussing first the limit $|Y|\ll1$, the scaling function $J_{\rm IA}(Y)$ is dominated by the divergent behavior of the momentum distribution $\tilde{n}(k)$ at small $k$. For generic Bose superfluids, this behaves like $\tilde{n}(k) = m c_s n_0/2n\hbar k$ at zero temperature~\cite{gavoret64} and like $\tilde{n}(k) = m^2 n_0 T/\rho_s \hbar^2 k^2$ at finite temperature~\cite{hohenberg65}, 
where $\rho_s$ is the superfluid (mass) density. As a result, the singular part of $J_{\rm IA}(Y)$ is given by
\begin{align}
&\lim_{|Y|\to 0} J_{\rm IA}(Y) \nonumber \\
&=n_0 \tilde{\xi}^3 \delta(Y) -  \frac{m \tilde{\xi}^2 n_0}{4 \pi^2 \hbar} \begin{cases} \frac{c_s}{2n} \frac{|Y|}{\tilde{\xi}} & T=0 \\ \frac{m T}{\hbar \rho_s} \ln |Y| & T\neq 0\end{cases} + {\rm const}, \label{eq:JIAsmallY}
\end{align}
i.e., a cusp at zero temperature and a logarithmic divergence $\sim n_0T\, \ln(1/|Y|)$  at finite temperature. In the opposite limit $|Y|\gg 1$, the scaling function $J_{\rm IA}(Y)$ depends on the behavior of the momentum distribution at large momenta, which is generically not universal. In the particular case of ultracold gases, however, the momentum distribution exhibits a universal power-law decay $n(k) = \mathcal{C}_2/k^4$ determined by the two-body contact density $\mathcal{C}_2$. For ultracold atoms, therefore, the scaling function $J_{\rm IA}(Y)$ for large values $|Y|\gg 1$ acquires a universal form 
\begin{align}
\lim_{|Y|\gg 1} J_{\rm IA}(Y) &= \frac{\tilde{\xi}^4 \mathcal{C}_2}{8 \pi^2Y^2} . \label{eq:JIAasym}
\end{align}
As was shown by Tan and by Braaten, Kang, and Platter~\cite{tan08b,braaten08,braaten11}, the high-momentum tail of the momentum distribution applies for arbitrary states of  either Bose or Fermi gases with zero-range interactions, both at zero temperature and in the nondegenerate limit, where it holds for wave vectors large compared to the inverse thermal length $\lambda_T$. Hence, while the small-$|Y|$ form~[Eq.~\eqref{eq:JIAsmallY}] is specific to Bose-condensed systems, the large-$|Y|$ tail~[Eq.~\eqref{eq:JIAasym}] is completely universal.

The IA does not take into account interactions between the scattered state and the initial state. As a result, it carries information about the time-dependent density correlations only through the equal-time momentum distribution. Corrections to the IA scaling form are suppressed as $\mathcal{O}(1/qa)$. Following the ground-breaking work of Hohenberg and Platzman, a number of attempts have been made to include interactions beyond the IA in a systematic expansion in inverse powers of momentum~\cite{gersch72,platzman84,carraro91}.  The terms in this expansion, however, involve the complete two-body and higher-body density matrices, which are not known in general. As we discuss above, the IA fails to account for processes where the probe scatters off pairs of high-momentum states or processes where interactions distribute the imparted large momentum between two or more atoms (cf. Fig.~\ref{fig:regimes}). By energy and momentum conservation, such processes become relevant if $\hbar \omega - \varepsilon_{\bf q} = \mathcal{O}(q^2)$, i.e., if $Y = \mathcal{O}(q)$. In the following, we show that,  at least for ultracold gases, this multiparticle regime can be described accurately by the OPE, i.e., the same method that is used in high-energy physics to account for the QCD interaction corrections to the parton model. The associated leading term in an expansion in inverse powers of momentum is given by Eq.~\eqref{eq:OPEscaling}, which involves only the two-body contact density.

Formally, the OPE expresses the product of two operators (which in the case of interest are the density operators) at different points in space and  time as a sum of local operators~\cite{braaten08,braaten12}:
\begin{align}
i \hat{n}(t,{\bf r}) \hat{n}(0,{\bf 0}) &= \sum_{\ell} W_\ell(t,{\bf r},a) \hat{ O}_\ell(0,{\bf 0}) . \label{eq:defOPE}
\end{align}
The dependence on the difference of the operator arguments is carried by the coefficients of this expansion $W_\ell(t,{\bf r},a)$ -- called the Wilson coefficients --  which are pure functions and not operators. This non-relativistic OPE is, in fact -- at least for special cases -- a convergent expansion~\cite{goldberger15}. Importantly, Eq.~\eqref{eq:defOPE} is an operator relation; i.e., it holds if we take its expectation value between arbitrary states. Using the OPE in Eq.~\eqref{eq:defchi} and performing the Fourier transformation gives
\begin{align}
&\chi(\omega, {\bf q}) = \sum_\ell \frac{m}{\hbar^2 q^{\Delta_\ell - 1}} J_\ell\Bigl(Z, \frac{1}{qa}\Bigr) \langle \hat{O}_\ell \rangle , \label{eq:OPEchi}
\end{align}
where we separate the $q$ dependence from the Wilson coefficient and write its remainder in terms of a dimensionless scaling function $J_\ell$ that depends on $(qa)^{-1}$ and $Z = \hbar \omega/\varepsilon_{\bf q} - 1$. The exponent of $\Delta_\ell - 1$ in front depends on the scaling dimension of the operators $\hat{O}_\ell$, which are formally defined through 
\begin{align}
\langle \hat{O}_\ell^\dagger(t, {\bf r}) \hat{O}_\ell(0,{\bf 0}) \rangle \sim \frac{1}{t^{\Delta_\ell}} \exp\biggl[- i N_\ell \frac{m r^2}{2 \hbar t}\biggr] ,
\end{align}
where $N_\ell$ denotes the number of particle creation or annihilation operators in $\hat{O}_\ell$. Since the scaling dimension in nonrelativistic theories is bounded from below~\cite{nishida12b}, the leading-order asymptotic form of the density response is determined by the operators with the lowest scaling dimension. Some details of the OPE calculation for the density response are given in Appendix~\ref{app:ope}. 
%++++++++++++++++++++++++++++++++++++++++
\begin{figure}[t]
\scalebox{1.}{\includegraphics{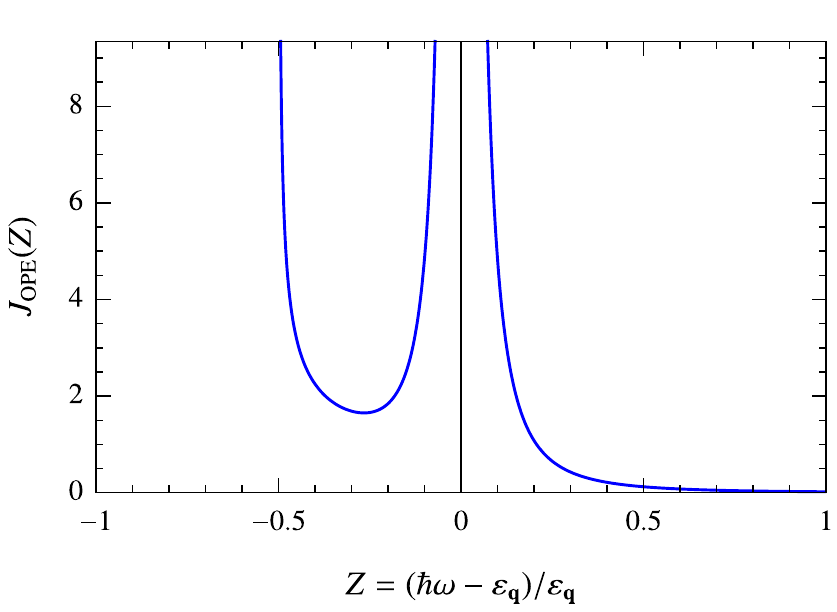}}
\caption{Scaling function $J_{\rm OPE}(Z)$ as given in Eq.~\eqref{eq:opedynstruc} as a function of the scaling variable $Z=(\hbar \omega - \varepsilon_{\bf q})/\varepsilon_{\bf q}$. The scaling function is symmetric near $Z=0$ but is strongly asymmetric for $|Z|=\mathcal{O}(1)$.}
\label{fig:JOPE}
\end{figure}
%++++++++++++++++++++++++++++++++++++++++
The leading-order term in Eq.~\eqref{eq:OPEchi} is set by the density operator ${O}_n$ with Wilson coefficient~\cite{son10,hofmann11,goldberger12, nishida12}
\begin{align}
J_n(\omega, {\bf q}) &= - \frac{2}{Z} + \frac{2}{Z+2} . \label{eq:Jn}
\end{align}
Note that at this leading level, the Wilson coefficient is independent of $qa$. For positive frequency, $Z > -1$, Eq.~\eqref{eq:Jn} gives rise to a delta peak at $\hbar \omega = \varepsilon_{\bf q}$ in the dynamic structure factor with weight $n$. It is important to note that this delta peak has nothing to do with the presence of a delta peak due to a nonvanishing condensate density $n_0$, as predicted by the IA. It merely reflects the fact that the OPE presents only a ``coarse-grained'' picture (as we discuss below) of the dynamic structure factor near the single-particle peak. The asymptotic form of the incoherent part away from $\hbar \omega = \varepsilon_{\bf q}$ is determined by the next-to-leading-order term in the OPE, which is set by the Wilson coefficient of the contact operator $\hat{O}_c$ with expectation value $\mathcal{C}_2 = \langle \hat{O}_c \rangle$. In order to to make contact with the IA, we consider in the following the limit $q|a|\gg1$.  In this limit, the leading contribution to the dynamic structure factor away from the single-particle peak is given by~\cite{son10,hofmann11,goldberger12,nishida12}
\begin{align}
&J_{\rm OPE}(Z) = \frac{1}{\pi} {\rm Im} \, J_\mathcal{C}\bigl[Z, (qa)^{-1} = 0\bigr] \nonumber \\
&= \frac{1}{2\pi^2} \biggl[\frac{\sqrt{2Z+1}}{Z^2} + \frac{1}{Z+1} \ln \frac{Z + 1 + \sqrt{2Z+1}}{|Z|} \nonumber \\
&\quad - \frac{2}{\sqrt{2Z+1}} \biggl(\ln^2 \frac{Z + 1 + \sqrt{2Z+1}}{|Z|} - \pi^2 \Theta(-Z)\biggr)\biggr] .\label{eq:opedynstruc}
\end{align}
This function is shown in Fig.~\ref{fig:JOPE}.  In contrast to the IA, it predicts a spectrum that is not symmetric around the single-particle peak at $Z=0$. In particular, it involves an onset singularity at $Z=-1/2$ and a power-law tail at high frequencies. In the following, we discuss the physics behind these features in detail, starting with the behavior near the single-particle peak, where the OPE turns out to be smoothly connected to the impulse approximation.

According to the OPE, the spectrum near the single-particle energy $\varepsilon_{\bf q}$ consists of a delta function of weight $n$ associated with the leading contribution~[Eq.~\eqref{eq:Jn}] and a singular background proportional to $1/Z^2$. This is quite different from the prediction of the IA, which involves a delta peak at $\hbar\omega=\varepsilon_{\bf q}$ whose weight is determined by the {\it condensate} density $n_0$ plus --- at $T=0$ --- a smooth, symmetric background. Now, in deriving the OPE, we rely on $\omega$ and $q$ being large compared to any other scale in the system. Indeed, when computing the Wilson coefficients by matching few-body matrix elements, all intrinsic energy and length scales are neglected; i.e., we drop any correction of order $\mathcal{O}(q^{-1})$. However, for energies $|\hbar \omega - \varepsilon_{\bf q}| \sim \mathcal{O}(q)$ close to the single-particle energy, there are contributions to the density response that probe {\it low-energy} properties of the gas even if $\omega$ and $q$ are large, such as processes where a large momentum is transferred to a single atom with small momentum. The behavior close to the single-particle peak can therefore {\it not} be resolved by the OPE. Remarkably, however, the OPE and IA can be smoothly connected near the crossover scale, where $Z=2Y/q\tilde{\xi}=\mathcal{O}(1/q)$, i.e., $Y=\mathcal{O}(1)$. To see this, consider the OPE scaling function~\cite{son10,hofmann11,goldberger12, nishida12},
\begin{align}
\lim_{Z\to 0} J_{\rm OPE}(Z, qa) &= \frac{1}{2 \pi^2 Z^2} \biggl[1 + \frac{2}{1+(qa/2)^2}\biggr] + \cdots , \label{eq:limitOPEBose}
\end{align}
near $Z=0$ for arbitrary values of the scattering length. Here, the first term in the square brackets coincides with the $1/qa=0$ result from Eq.~\eqref{eq:opedynstruc}. Comparing with the few-particle calculations of Refs.~\cite{son10,hofmann11,goldberger12, nishida12} that determine the Wilson coefficients, we can interpret the first term as a self-energy correction to the initial or final state, and the remaining term --- which depends on the scaling variable $qa$ --- as a final-state vertex correction.

We now make the following very important observation: in the high-momentum limit $qa \gg 1$, the term in Eq.~\eqref{eq:limitOPEBose} that we recognize as a vertex correction vanishes near the single-particle peak. The OPE result ,
\begin{align}
S_{\rm OPE}(\omega,\mathbf{q})\to \frac{m\mathcal{C}_2}{\hbar^2q^3} \frac{1}{2\pi^2Z^2}=\frac{m}{\hbar^2\tilde{\xi}^2\, q} \frac{\tilde{\xi}^4\mathcal{C}_2}{8\pi^2Y^2} , \label{eq:OPE=IA}
\end{align} 
thus coincides with the $|Y|\gg 1$ limit of the IA as determined by Eq.~\eqref{eq:JIAasym}.  For large momentum $qa\gg 1$, the IA and OPE are therefore complementary scaling functions that describe separate asymptotic high-momentum regimes. They match smoothly in the regime where $|Y|\gg 1$ and, thus, where $|Z|=\mathcal{O}(1/q\tilde{\xi})$ is small. This scaling behavior is sketched in Fig.~\ref{fig:scaling}. Away from unitarity [i.e., for $(qa)^{-1} = \mathcal{O}(1)$], the small-energy deviations are no longer described by the IA, and vertex corrections need to be taken into account. These corrections due to a finite scattering length have been calculated in Ref.~\cite{hofmann11}, and we make use of these results in Sec.~\ref{sec:lineshift} when discussing the line shift of the single-particle peak for arbitrary values of $qa$.

As a second point, we discuss the origin of the sharp onset of the scaling function $J_{\rm OPE}(Z)$ at $Z = -1/2$. This left boundary is of kinematic origin and marks the minimum energy $\hbar\omega$ that a probe with fixed large wave number $q$ can impart on two atoms at rest. Note that  the threshold for multiparticle excitations lies below the position of the single-particle peak, quite different from what happens in the long-wavelength limit. The behavior of the dynamic structure factor near the two-particle threshold is dictated by the form of the two-particle $T$ matrix~\cite{son10,hofmann11,goldberger12,nishida12}. In the special case of infinite scattering length $(qa)^{-1} = 0$, which is considered in Fig.~\ref{fig:JOPE}, the dynamic structure factor at the two-particle threshold diverges as $1/\sqrt{\hbar \omega - \varepsilon_{\bf q}/2}$. For finite scattering length, in turn, this divergence disappears and the structure factor vanishes according to the Wigner threshold law $\sqrt{\hbar \omega - \varepsilon_{\bf q}/2}$~\cite{hofmann11}.

Concerning the behavior in the deep inelastic limit $Z\gg 1$ far to the right of the single-particle peak, the OPE scaling function falls off as $Z^{-7/2}$.  This is a special case of the more general result~\cite{son10,hofmann11,goldberger12, nishida12,taylor10}
\begin{align}
S(\omega, {\bf q}) = \frac{\hbar^{3/2} q^4 \mathcal{C}_2}{m^{5/2} \pi^2 \omega^{7/2}} \biggl[\frac{2}{45} + \frac{1}{72} \frac{(a^{-1}/\sqrt{m\omega/\hbar})^2}{(a^{-1}/\sqrt{m\omega/\hbar})^2 + 1}\biggr] \label{eq:tail}
\end{align}
for the high-frequency tail of the dynamic structure at arbitrary values of the scattering length. The physics underlying this tail was discussed some time ago by Wong~\cite{wong77}: it is due to two-particle excitations, which -- at large wave vectors -- have energy $2\varepsilon_{\bf q}$. The incoherent part of the dynamic structure factor,
 \begin{align}
S_{\rm inc}(\omega, {\bf q}) &=\frac{1}{(\hbar\omega)^4} \int \frac{d^3k}{(2\pi)^3} \delta(\hbar\omega-\varepsilon_{\bf k}-\varepsilon_{\bf q-k}) \nonumber \\
&\qquad \times \bigl|\bigl\langle {\bf k}, {\bf q}-{\bf k} \bigr| [H, [H, \hat{\rho}_{\bf q}^\dagger]] \bigl| 0 \bigr\rangle \bigr|^2 ,
\end{align} 
is calculated by using to leading order in ${\bf q}$ the double commutator  
\begin{align}
[H, [H, \hat{\rho}_{\bf q}^\dagger]] = \frac{\hbar^2}{2 m} \int d({\bf x}, {\bf y}) ({\bf q} \cdot \hat{\bf r})^2 [r V'(r)] \hat{\rho}({\bf x}) \hat{\rho}({\bf y}) +\cdots ,
\end{align} 
which can be expressed as a product of two density operators for any general, spherically symmetric interaction potential $V(r)$.  In the particular case of a zero-range pseudopotential $V({\bf r}) = \frac{4\pi\hbar^2a}{m} \delta({\bf r})$ and for a Bose-condensed system, where $\hat{\rho}_{\bf k} = \sqrt{N_0} (a_{\bf k}^\dagger + a_{-{\bf k}})$ to leading order, this gives $S_{\rm inc}(\omega, {\bf q})  =  \frac{\hbar^2 q^4}{32 \pi^2 m^3\omega^4} \sqrt{m\omega/\hbar} (4\pi n_0a)^2$ by using free two-particle states $|{\bf k}, {\bf q}-{\bf k}\rangle = a_{\bf k}^\dagger a_{\bf q-k}^\dagger |0\rangle$~\cite{wong77}.  This has the same form as the exact OPE result in Eq.~\eqref{eq:tail} and --- in particular --- it gives the correct $q^4/\omega^{7/2}$ scaling with a contact density $ \mathcal{C}_2\to (4\pi n_0a)^2$.  Because of the simple two-particle ansatz, which neglects all final-state interactions, the result, however, fails to capture both the correct prefactor and the general expression for the contact density, which is finite even without any condensate. Remarkably, the high-frequency tail appears consistent with $n$-scattering data on $^4$He at $T=1.2\,$K and a large wave vector $q=0.8\,\AA^{-1}$ in a restricted range of energies $25 \,K<\hbar\omega/k_B<70\, K$~\cite{wong77}, despite the fact that the interactions between helium atoms are quite different from the zero-range interactions present in ultracold gases.

Finally, we briefly comment on the effect of operators with higher scaling dimension. If three-particle and higher-order excitations are taken into account, the onset of the incoherent weight of the dynamic structure factor shifts to even lower frequencies. As noted by Son and Thompson~\cite{son10}, there is a cascade of threshold frequencies $\hbar \omega_n = \varepsilon_{\bf q}/n$ above which $n$-body excitations contribute. Because of their higher scaling dimensions, they are suppressed at high momentum according to Eq.~\eqref{eq:OPEchi} compared to excitations involving fewer particles. Nevertheless, $n$-body excitations dominate the dynamic structure factor in an energy interval $\varepsilon_{\bf q}/n \leq \hbar \omega \leq \varepsilon_{\bf q}/(n-1)$. Specifically, the scaling near the $n$-body threshold in the absence of any fine-tuning of the scattering length is given by $(\hbar \omega - \varepsilon_{\bf q}/n)^{(3n-5)/2}$, in accordance with the Wigner threshold law~\cite{sadeghpour00}. Moreover, the high-frequency tail to the right of the single-particle peak decays as a power law as $q^4/\omega^{(\Delta_{\mathcal{C}_n}+3)/2}$, where $\Delta_{\mathcal{C}_n}$ denotes the scaling dimension of the $n$-body contact parameter. The leading term beyond the contribution from two-particle correlations, which are described by Eq.~\eqref{eq:OPEscaling}, involves three particles. The associated contribution to the dynamic structure factor is proportional to the so-called three-body contact $\mathcal{C}_3$, which may be defined by the dependence~\cite{braaten11}
\begin{align}
\mathcal{C}_3 = - \frac{m\kappa_*}{2} \frac{\partial \mathcal{E}}{\partial \kappa^*} \biggr|_{a^{-1}}
\end{align}
of the energy density $\mathcal{E}$ on the three-body parameter $\kappa^*$, which is necessary as a short-distance cutoff to stabilize a Bose gas with zero-range interactions. Most notably, the three-body contact sets the magnitude of the subleading correction to the momentum distribution, which is predicted to decay as~\cite{braaten11}
\begin{align}
n(k) = \frac{\mathcal{C}_2}{k^4} + \frac{\mathcal{C}_3}{k^5} F(k)+\cdots \, .  \label{eq:tailC3}
\end{align}
Here, $F(k) = A \sin(2s_0 \ln \frac{k}{\kappa_*} + 2 \phi)$ is a log-periodic function that depends on the value of the three-body parameter, while $s_0 = 1.00624$, $\phi = - 0.669064$, and $A = 89.26260$ are universal numerical constants. Near the three-body threshold, the dynamic structure factor vanishes like $S(\omega,{\bf q}) \sim (\hbar \omega - \varepsilon_{\bf q}/3)^2 \mathcal{C}_3$, according to the Wigner threshold law for $n=3$. The Bose gas has a renormalization group limit cycle in the three-particle sector, which is caused by the Efimov effect. As a result, the scaling dimension $\Delta_{\mathcal{C}_3} = 5 + 2 i s_0$ has a nonvanishing imaginary part that is determined by the universal Efimov number  $s_0$~\cite{nishida12b}. This implies a high-frequency tail of the form
\begin{align}
S(\omega, {\bf q}) \to \frac{2 \hbar^{3/2}}{45 \pi^2 m^{5/2}} \frac{q^4 \mathcal{C}_2}{\omega^{7/2}} + \frac{A q^4 \mathcal{C}_3}{\omega^4} \sin(s_0 \ln \omega/B)\, , \label{eq:OPEC3}
\end{align}
where $A$ and $B$ are constants.

\section{Sum rules at large momentum transfer}\label{sec:sumrules}

In the previous section, we obtain an expression for the dynamic structure factor valid at high momentum which covers the full range of frequencies. As an important check of our results, in the following we compute various sum rules for which exact results are known.

\subsection{Moment sum rules}

The moment sum rules are defined as
\begin{align}
m_p &= \hbar^{p+1} \int_{-\infty}^\infty d\omega \, \omega^p S(\omega,{\bf q}) . \label{eq:defmp}
\end{align}
For $p=-1$, we obtain the compressibility sum rule. By the Kramers-Kronig relation, it is related to the static limit of the dynamic density response function $\chi(\omega = 0,{\bf q})$~\cite{pitaevskii16}, which at high momentum can be inferred from the results presented in Refs.~\cite{goldberger12,hofmann11}:
\begin{align}
m_{-1} &= \frac{1}{2} \chi(\omega = 0,{\bf q}) \nonumber \\
&\stackrel{q\to \infty}{\to} \frac{n}{\varepsilon_{\bf q}} + \frac{\pi \mathcal{C}_2}{8 \varepsilon_{\bf q} q} \biggl( 1 - \frac{8 + 24 \pi - 6 \pi^2}{3\pi^2 q a} \biggr) + \cdots , \label{eq:mn1}
\end{align}
where corrections arise at ${\cal O}(q^{-5})$ and from operators with higher scaling dimension. This high-momentum form of the compressibility sum rule is a new result. The zeroth moment $m_0$ defines the static structure factor, which at high momentum reads
\begin{align}
m_0 &= n S(q) \stackrel{q\to \infty}{\to} n\biggl(1 + \frac{\mathcal{C}_2}{8 n q} \biggl[1 - \frac{4}{\pi qa}\biggr] + \cdots \biggr) . \label{eq:Sq}
\end{align}
It is instructive to compare this exact result with the behavior obtained in the Bogoliubov approximation where, as pointed out above,  the single-mode approximation is exact at {\it arbitrary} momenta.  As a result, for a weakly interacting Bose gas, one has $S({\bf q}) = \varepsilon_{\bf q}/E_{\bf q}$, with $E_{\bf q} = \sqrt{\varepsilon_{\bf q} (\varepsilon_{\bf q} + 2 gn)}$ being the Bogoliubov energy and $g=4\pi\hbar^2a/m$. For large momenta $q\xi\gg 1$, the static structure factor thus approaches its trivial limit of unity like $S({\bf q}) \to 1 - 1/(q\xi)^2$, missing the positive $\mathcal{C}_2/q$ part.  As we discuss above, Bogoliubov theory does not account for the positive $\mathcal{C}_2/(4\pi nr)^2$ contribution to the pair -distribution function which gives rise to the leading $\mathcal{C}_2/(8nq)$ term in the high-momentum limit of the static structure factor 
\begin{align}
S(\mathbf{q})=1+n \int\! d{\bf r} \, e^{-i\mathbf{q}\cdot\mathbf{r}}\, \bigl[g^{(2)}(\mathbf{r})-1\bigr]\, . \label{eq:defSq}
\end{align}
As we discuss below, this positive contribution, which becomes appreciable for momenta $qa=\mathcal{O}(1)$, gives rise to a maximum in the static structure factor, providing a qualitative explanation of the nonmonotonic behavior of the level shift observed in Ref.~\cite{papp08}.

The first moment $m_1$ in Eq.~\eqref{eq:defmp} is the $f$-sum rule, Eq.~\eqref{eq:fsumrule}, which is unaffected by interactions as long as no velocity-dependent contributions are present. The second moment sum rule is sensitive to the total kinetic energy and is, hence, known as the kinetic sum rule~\cite{puff65,stringari92}. Because of the high-frequency tail $\sim\omega^{-7/2}$ we discuss in Eq.~\eqref{eq:tail}, which holds for arbitrary values of the scattering length, the third and higher moments are no longer finite.

In order to verify the results for the dynamic structure factor at large momentum derived in Sec.~II  in a completely different manner, we compute in the following the contributions of the IA and OPE to the moments at large momentum transfer by splitting the frequency integration in two regions where either the IA or the OPE applies:
\begin{align}
m_p &\stackrel{q\to \infty}{\to} m_p^{({\rm IA})}(\eta) + m_p^{({\rm OPE})}(\eta) + \cdots .
\end{align}
Corrections to this decomposition only appear at smaller momentum transfer where the asymptotic form of $S(\omega,{\bf q})$ is no longer given by the combination of the OPE and IA. For the first part, $m_p^{({\rm IA})}(\eta)$, we restrict the frequency integration in Eq.~\eqref{eq:defmp} to the vicinity of the single-particle peak $\varepsilon_{\bf q} - \eta \leq \hbar \omega \leq \varepsilon_{\bf q} + \eta$. Here, the energy scale $\eta$ is chosen in such a way that $1/\tilde{\xi} q \ll \eta/\varepsilon_{\bf q} \ll 1$. This limit marks the crossover region between IA and OPE, where the dynamic structure factor is given by Eq.~\eqref{eq:opedynstruc}. In the remaining integration region, which defines $m_p^{({\rm OPE})}(\eta)$, the OPE result applies. While both contributions depend explicitly on $\eta$, this dependence cancels when adding both contributions. The individual contributions $m_p^{({\rm IA})}(\eta)$ and $m_p^{({\rm OPE})}(\eta)$ could be useful as restricted sum rules that apply to the $Y$- and $Z$-scaling regime.

The IA contribution is given by
\begin{align}
m_p^{({\rm IA})} &= \frac{1}{4\pi^2} \int_0^\infty dk \, k^2 n(k) \int_{-1}^1 dx \, \nonumber \\
&\qquad\qquad\times \biggl(\varepsilon_{\bf q} + \frac{\hbar^2}{m} {\bf k} \cdot {\bf q}\biggr)^p \, \Theta\Bigl(\Bigl|\frac{\hbar^2}{m} {\bf k} \cdot {\bf q}\Bigr| \leq \eta\Bigr) .
\end{align}
The angle integration can be performed in closed analytical form. The remaining momentum integration is carried out using the high-momentum tail of the momentum distribution $n(k) = \mathcal{C}_2/k^4$, yet without imposing an explicit form of the momentum distribution. The result is
\begin{gather}
m_{-1}^{({\rm IA})} = \frac{n}{\varepsilon_{\bf q}} - \frac{\hbar^2 q}{4 \pi^2 m} \frac{\mathcal{C}_2}{\varepsilon_{\bf q} \eta} \\
m_{0}^{({\rm IA})} = n - \frac{\hbar^2 q}{4 \pi^2 m} \frac{\mathcal{C}_2}{\eta} \\
m_{1}^{({\rm IA})} = n \varepsilon_{\bf q} - \frac{\hbar^2 q}{4 \pi^2 m} \frac{\varepsilon_{\bf q} \mathcal{C}_2}{\eta} \\
m_{2}^{({\rm IA})} = n \varepsilon_{\bf q}^2 + \frac{4 \varepsilon_{\bf q}}{3} \mathcal{E} - \frac{\hbar^2 q}{4 \pi^2 m} \frac{\varepsilon_{\bf q}^2 \mathcal{C}_2}{\eta} .
\end{gather}
Here, $\mathcal{E} = \int_{\bf k} \varepsilon_{\bf k}\bigl[n(k) - \mathcal{C}_2/k^4\bigr]$ is the energy density of the unitary Bose gas~\cite{schakel10,braaten11}. It is important to note that these expressions are not affected by the presence of a condensate peak in the momentum distribution. The result for the moments is, thus, not restricted to the Bose-condensed phase and holds equally well in the normal phase.

The OPE contribution to the sum rules are obtained by a direct calculation using Eq.~\eqref{eq:opedynstruc}. We obtain
\begin{gather}
m_{-1}^{({\rm OPE})} = \frac{\pi \mathcal{C}_2}{8 \varepsilon_{\bf q} q} + \frac{\hbar^2 q}{4 \pi^2 m} \frac{\mathcal{C}_2}{\varepsilon_{\bf q} \eta} \\
m_{0}^{({\rm OPE})} = \frac{\mathcal{C}_2}{8 q} + \frac{\hbar^2 q}{4 \pi^2 m} \frac{\mathcal{C}_2}{\eta} \\
m_{1}^{({\rm OPE})} = \frac{\hbar^2 q}{4 \pi^2 m} \frac{\varepsilon_{\bf q} \mathcal{C}_2}{\eta} \\
m_{2}^{({\rm OPE})} = \frac{\hbar^2 q}{4 \pi^2 m} \frac{\varepsilon_{\bf q}^2 \mathcal{C}_2}{\eta} .
\end{gather}
As expected, the $\eta$ dependence cancels when summing the two contributions. The final results are in agreement with the general results obtained in the previous section. The result for the second moment at large $qa$,
\begin{align}
m_{2} &= n \varepsilon_{\bf q}^2 + \frac{4 \varepsilon_{\bf q}}{3} \mathcal{E} , \label{eq:m2}
\end{align}
is new, however.

\subsection{Borel sum rule}

As a generalization of the moment sum rules, Goldberger and Rothstein~\cite{goldberger12} consider a so-called Borel sum rule with an exponential weight factor defined by
\begin{align}
m_B = \frac{1}{\omega_0^2} \int_{0}^\infty d\omega \, \omega \, e^{-\omega^2/\omega_0^2} S(\omega,{\bf q}) .\label{eq:borel}
\end{align}
In a high-energy context, sum rules of this type are used to constrain hadronic properties~\cite{shifman79a,shifman79b}, and they have recently also been used to constrain the spectral function of a quantum gas~\cite{gubler15}. The sum rule contains an arbitrary weight parameter $\omega_0$ such that the contribution of the high-frequency part of $S(\omega,{\bf q})$ to the sum rule becomes more dominant with increasing values of $\omega_0$. In particular, the OPE can be used to compute the sum rule in an expansion in the small parameter $\varepsilon_{\bf q}/(\hbar\omega_0)$~\cite{goldberger12}\footnote{Note that this result differs from Eq.~(90) of Ref.~\cite{goldberger12} in the sign of the $\mathcal{E}$ term. We also include a higher-order correction in the density, which is of the same order $\mathcal{O}(\omega_0^{-4})$ as the contribution of the contact.}:
\begin{align}
m_B &= \frac{n \varepsilon_{\bf q}}{(\hbar\omega_0)^2} - \frac{n \varepsilon_{\bf q}^3}{(\hbar\omega_0)^4} - \frac{4 \varepsilon_{\bf q}^2}{(\hbar \omega_0)^4} \mathcal{E} \nonumber \\
& - \frac{16\sqrt{2}\hbar \varepsilon_{\bf q}^2}{135\pi m^{1/2}\Gamma(3/4)} \frac{\mathcal{C}_2}{(\hbar \omega_0)^{7/2}} + \mathcal{O}\Bigl(\frac{\varepsilon_{\bf q}^3}{(\hbar \omega_0)^3}\Bigr) . \label{eq:borelexact}
\end{align}
As in the previous section, this result for the sum rule also follows from the exact asymptotic form of the dynamic structure factor, and restricted sum rules valid in the regime of IA and OPE, respectively, can be derived. The IA contribution to the Borel sum rule is
\begin{align}
&m_B^{({\rm IA})} = \frac{n\varepsilon_{\bf q}}{(\hbar\omega_0)^2} - \frac{n \varepsilon_{\bf q}^3}{(\hbar\omega_0)^4} - \frac{4 \varepsilon_{\bf q}^2 \mathcal{E}}{(\hbar\omega_0)^4} \nonumber \\
&- \frac{\hbar^2 \varepsilon_{\bf q}  q}{4\pi^2m\eta} \frac{\mathcal{C}_2}{(\hbar\omega_0)^2} + \frac{\hbar^2 \varepsilon_{\bf q}^3 q}{4 \pi ^2 m \eta} \frac{\mathcal{C}_2}{(\hbar \omega_0)^4} + \mathcal{O}\biggl(\frac{\varepsilon_{\bf q}^6}{(\hbar \omega_0)^6},\frac{\eta}{\varepsilon_{\bf q}}\biggr) .
\end{align}
The OPE contribution is
\begin{align}
m_B^{({\rm OPE})} &= \frac{\hbar^2 \varepsilon_{\bf q} q}{4 \pi^2 m \eta} \frac{\mathcal{C}_2}{(\hbar\omega_0)^2} - \frac{\hbar^2 \varepsilon_{\bf q}^3 q}{4\pi^2 m \eta} \frac{\mathcal{C}_2}{(\hbar\omega_0)^4} \nonumber \\
&- \frac{16\sqrt{2}\hbar \varepsilon_{\bf q}^2}{135\pi m^{1/2}\Gamma(3/4)} \frac{\mathcal{C}_2}{(\hbar \omega_0)^{7/2}} + \mathcal{O}\biggl(\frac{\varepsilon_{\bf q}^6}{(\hbar \omega_0)^6},\frac{\eta}{\varepsilon_{\bf q}}\biggr) .
\end{align}
The sum of these two contributions gives the full Borel sum rule~\eqref{eq:borel} in agreement with Eq.~\eqref{eq:borelexact}, free of any $\eta$ dependence.

\section{line shift}\label{sec:lineshift}

In the experiments by Papp {\it et al.}~\cite{papp08}, the Feshbach resonance in $^{85}$Rb near $B_0=155\,$ G is used to increase the scattering length to values up to $10^3\, a_0$. For a fixed wave vector of the Bragg pulse, this gives access to the dynamic structure factor in a regime where $q\xi$ is much larger than one and the momentum transfer $q$ is also of order of or larger than the inverse scattering length $1/a$. For a fixed wave vector of the Bragg pulse, this gives access to the dynamic structure factor in a regime where the momentum transfer $q$ is of order of or larger than the inverse coherence length $1/\xi$, with typical values $q\xi\simeq 2-3$. The peak position, which at large momentum will eventually be centered right at the single-particle energy $\varepsilon_{\bf q}$, has a correction due to interactions which defines the line shift. While Bogoliubov theory predicts a linear line shift at small scattering length,
\begin{align}
\Delta (\hbar \omega) = \hbar \omega_{\bf q} - \varepsilon_{\bf q} &= \frac{4 \pi \hbar^2 a n}{m}  + \mathcal{O}\Bigl(\frac{1}{q^2}\Bigr) , \label{eq:bogoliubovshift}
\end{align}
the measurement~\cite{papp08} starts off linearly but then shows a downturn with increasing scattering length once $qa\simeq 1$. In this section, we use the operator product expansion to compute the line shift at high momentum $q\xi\gg 1$ allowing, however, for arbitrary values of $qa$. While the previous sections are concerned with the fine structure near the single-particle peak at $qa\to \infty$, here, we are interested in the broad structure of the peak for all $qa$, i.e., its position and width. For these quantities, we can apply the OPE to obtain universal results for line shift and width that depend on the Tan two-body contact parameter $\mathcal{C}_2$.

We begin by considering the structure of the density response near the single-particle peak, which takes the general form
\begin{align}
\chi(\omega, {\bf q}) &= - \frac{Z_{\bf q}}{\hbar \omega - \varepsilon_{\bf q} - \Pi(\omega, {\bf q})} + \chi^{\rm inc}(\omega, {\bf q}) ,
\end{align}
where $\chi^{\rm inc}$ denotes the incoherent part. The position of the one-particle peak is defined by the zeros of the denominator
\begin{align}
\hbar \omega - \varepsilon_{\bf q} - {\rm Re} \, \Pi(\omega, {\bf q}) &= 0 
\end{align}
 at $\omega=\omega_{\bf q}$. The imaginary part at the resonance frequency $\omega_{\bf q}$ determines the width $\Gamma$ of the peak as $\Gamma = - {\rm Im} \, \Pi(\omega_{\bf q}, {\bf q})$. At large momentum, the many-body correction induced by $\Pi$ is subleading, and we can determine the new pole in the on-shell approximation
\begin{align}
\Delta (\hbar \omega) &= {\rm Re} \, \Pi(\varepsilon_{\bf q}, {\bf q}) + \mathcal{O}\Bigl(\frac{1}{\varepsilon_{\bf q}}\Bigr) .
\end{align}
Expanding the density response to leading order in $\Pi$,
\begin{align}
\chi(\omega, {\bf q}) &= - \frac{Z_{\bf q}}{\hbar \omega - \varepsilon_{\bf q}} -  \frac{Z_{\bf q} \Pi(\varepsilon_{\bf q}, {\bf q})}{(\hbar \omega - \varepsilon_{\bf q})^2} + \cdots ,
\end{align}
we infer the high-momentum structure of $Z_{\bf q}$ and $\Pi$ by comparing with the results of the operator product expansion~\cite{hofmann11,goldberger12}. This gives to leading order $Z_{\bf q} = n$ and
\begin{align}
\Pi(\varepsilon_{\bf q}, {\bf q}) &= \biggl[\frac{1}{2\pi a^2} \frac{1}{a^{-1} + i q/2} - \frac{iq}{8\pi} - \frac{1}{4\pi a}\biggr] \frac{\hbar^2 \mathcal{C}_2}{m n} . \label{eq:Pionshell}
\end{align}
This is one of the central results of this paper. The real part of $\Pi$ gives the line shift at large momentum transfer:
\begin{align}
\Delta (\hbar \omega) &\stackrel{q\to \infty}{\to}  \frac{\hbar^2 \mathcal{C}_2}{4 \pi m a n} \, \biggl[\frac{2}{1+ (qa/2)^2} - 1\biggr] . \label{eq:shift}
\end{align}
In addition, the imaginary part of Eq.~\eqref{eq:Pionshell} sets the width of the peak:
\begin{align}
\Gamma &\stackrel{q\to \infty}{\to}  \frac{\hbar^2 \mathcal{C}_2 q}{8 \pi m n} \, \biggl[\frac{2}{1+ (qa/2)^2} + 1\biggr] . \label{eq:width}
\end{align}
For weakly interacting gases with $n^{1/3}a\ll 1$, inserting the associated value $\mathcal{C}_2 = (4 \pi n a)^2$ of the contact density in Eq.~\eqref{eq:shift} reproduces the Bogoliubov result~[Eq.~\eqref{eq:bogoliubovshift}] at small $a$. As $a$ is increased, the prediction~[Eq.~\eqref{eq:shift}] deviates from Bogoliubov theory: it approaches a maximum at $qa \sim 1$, then bends backwards, and even changes its sign at large scattering length. Note that if the system is probed at wavelengths that are small compared to the interparticle distance $q \gg n^{1/3}$, the maximum may occur well in the perturbative region (because $n^{1/3} a \ll qa$).  At very large scattering length $qa \to \infty$, the line shift approaches zero from negative values as
\begin{align}
\lim_{a\to\infty} \Delta(\hbar \omega ) &\stackrel{q\to \infty}{\to}  - \frac{\hbar^2 \mathcal{C}_2}{4 \pi m a n} .
\end{align}

%++++++++++++++++++++++++++++++++++++++++
\begin{figure}[t]
\scalebox{1.}{\includegraphics{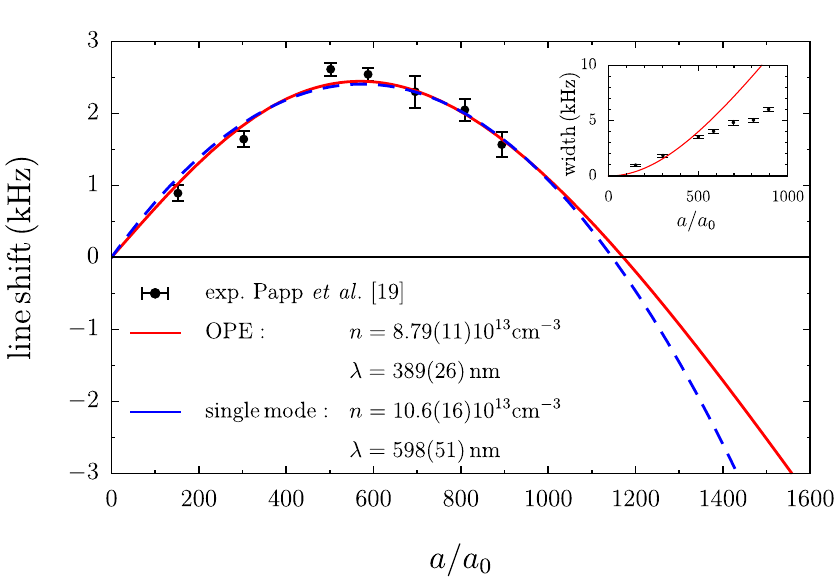}}
\caption{Line shift of a repulsive Bose gas as a function of scattering length $a$. The line shift~[Eq.~\eqref{eq:shift}] predicted by the operator product expansion is indicated by a red continuous line. For comparison, we include the line shift~[Eq.~\eqref{eq:singlemode}] as predicted from the single-mode ansatz (blue dashed line). Inset: OPE prediction~[Eq.~\eqref{eq:width}] for the width of the Bragg peak. The black points are the experimental results by Papp {\it et al.}~\cite{papp08}.}
\label{fig:line shift}
\end{figure}
%++++++++++++++++++++++++++++++++++++++++
 
Figure~\ref{fig:line shift} shows the comparison of the OPE prediction~[Eq.~\eqref{eq:shift}] (continuous red line) with the experimental results~\cite{papp08}. In our fit, we use the leading-order perturbative expression $\mathcal{C}_2 = (4\pi a n)^2$ for the contact parameter, since it turns out that for the scattering lengths in the experiment~\cite{papp08}, where $\sqrt{na^3} \sim \mathcal{O}(2-3)$, Lee-Huang-Yang and higher corrections to the contact are still rather small. For both fits, we fit an effective trap density $n$ and wavelength $\lambda$ of the Bragg beam (where $q=4\pi/\lambda$). Apparently,  the theory predictions in Fig.~\ref{fig:line shift} are in excellent agreement with the experimental data points.  Several caveats, however, apply to this comparison: (a) the experiment~\cite{papp08} probes the dynamic structure factor with $q\xi \simeq 2$ while our results here assume $q\xi\gg 1$; (b) we do not account for effects of a trap and perform a fit of the homogeneous result~[Eq.~\eqref{eq:shift}] with variable density and Bragg wavelength~\cite{zambelli00}. While the densities are in good agreement with the values quoted in Ref.~\cite{papp08}, the fitted Bragg wavelength is too small by a factor of almost $2$ compared to $\lambda = 780 \, {\rm nm}$ in~\cite{papp08}. Current experiments with $^{39}$K in box potentials, in fact, essentially eliminate trap effects and allow for a direct comparison of the line shift with the universal result~[Eq.~\eqref{eq:shift}] in a wide regime up to values $qa\simeq 8$~\cite{lopes17}. 

An often-used tool to estimate the collective mode frequencies of a quantum gas is the single-mode approximation~\cite{pitaevskii16}. It imposes the simple form $S^{\rm SM}(\omega, {\bf q}) = Z_{\bf q} \delta(\hbar \omega - \hbar \omega_{\bf q})$ discussed in the Introduction for {\it arbitrary} large values of the wave vector ${\bf q}$. The single-mode position $\hbar \omega_{\bf q}$ is then fixed by the ratio of two consecutive sum rules, such as $m_1/m_0$, $m_2/m_1$, or $m_0/m_{-1}$, which should all agree. In particular, assuming that the resonance peak in the dynamic structure factor is {\it below} the onset of an additional incoherent spectral weight, the single-mode approximation yields an upper bound on the true resonance position. Based on our exact results in Eqs.~\eqref{eq:fsumrule},~\eqref{eq:mn1},~\eqref{eq:Sq}, and~\eqref{eq:m2} for the sum rules at high momentum, the standard choice of the ratio $m_1/m_0$, for example, gives
\begin{align}
\hbar \omega_{\bf q}^{\rm SM} = \frac{m_1}{m_0} = \varepsilon_{\bf q} - \frac{\hbar^2 \mathcal{C}_2 q}{16 m n} \biggl[1 - \frac{4}{\pi qa}\biggr] + \cdots . \label{eq:singlemode}
\end{align}
We show the fit based on this form of the single-mode approximation as a blue dashed line in Fig.~\ref{fig:line shift}. Apparently, despite the fact that the underlying assumptions are not consistent with the exact results on the detailed spectrum obtained above,   Eq.~\eqref{eq:singlemode} agrees with~\eqref{eq:shift} to leading order in $qa\ll 1$. Moreover, it predicts a zero crossing of the line shift at $\bar{q}a=4/\pi$, which is of the same order as the prediction that $\Delta\equiv 0$ at $\bar{q}a=2$, which follows from the exact expression Eq.~\eqref{eq:shift}. It should be stressed, however, that the single-mode result~[Eq.~\eqref{eq:singlemode}] is larger than the exact result~[Eq.~\eqref{eq:shift}] for all $qa$ and, therefore, does not provide an upper bound. In addition, the agreement with Bogoliubov mean-field theory is specific to the ratio $m_1/m_0$, while for $m_0/m_{-1}$ this is no longer the case. Most importantly, the single-mode approximation does not account for a finite width of the peak, in stark contrast with the experimental results, which observe a substantial broadening of the Bragg peak with increasing values of the scattering length.

The significance of the result~[Eq.~\eqref{eq:shift}] for the line shift becomes clear when seen as a function of momentum: it predicts a negative  line shift at large momentum, which changes sign and becomes positive as the momentum is lowered beyond a critical value $\bar{q} a=2$. If such a behavior persisted to a region where Eq.~\eqref{eq:shift} is comparable to the single-particle energy $\varepsilon_{\bf q}$, the dispersion would no longer be monotonic but show a minimum at finite momentum, similar as for the roton in $^4$He. As we discuss in the Introduction, the Feynman single-mode ansatz~\cite{feynman54} -- which predicts that the position $\omega_{\bf q}$ of the peak in the dynamic structure factor is related to the static structure factor $S({\bf q})$ by $\hbar \omega_{\bf q} = \varepsilon_{\bf q}/S({\bf q})$ -- links the roton minimum to the nearest-neighbor pair correlations: in a quantum {\it liquid}, where the interparticle distance is comparable to the range $r_0$ of the interatomic interaction, $nr_0^3 \approx 1$, correlations over the size $r_0$ lead to a sharp peak in $S({\bf q})$ which causes a roton minimum in the single-particle dispersion. In a dilute quantum gas where $nr_0^3 \ll 1$, such a minimum is absent unless one considers in addition a strong dipolar contribution to the interaction~\cite{santos03,odell03,klawunn09}. Even for the case of zero-range interactions we consider here, however, a broad maximum is present in the high-momentum form of the static structure factor~[Eq.~\eqref{eq:Sq}] near the wave vector $\bar{q}a=2$ where the level shift changes sign. It appears at $\bar{q}a=8/\pi \approx 2.54$ (or $\bar{q}\xi \approx 0.51/\sqrt{na^3}$) and its value is \footnote{It is interesting to remark that even for $^4$He, the maximum value of the structure factor at $q_0$ is only around $1.4$, not far from the universal value for the transition to a quantum crystal, which appears if the maximum of $S(q)$ exceeds $1.4$ as a quantum generalization of the Hansen-Verlet criterion $S(q_0)\simeq 2.85$ for the freezing of classical fluids, see  M. H. Kalos, D. Levesque and L. Verlet, Phys. Rev. A {\bf 9}, 2178 (1974)}
\begin{align}
S(\bar{q}) = 1 + \frac{\pi a \mathcal{C}_2}{128 n} .
\end{align}
Quite generally, the maximum in the static structure factor of both strongly interacting Fermi and Bose gases in three dimensions is seen in full calculations with a position that is set by the interparticle distance~\cite{holzmann99,combescot06,hu10}. The presence of a maximum in the level shift~[Eq.~\eqref{eq:shift}] can thus be interpreted as a roton precursor in a dilute but strongly interacting quantum gas. It is interesting to contrast this with the behavior found for Bose gases in {\it one dimension}, where even in the limit of infinite zero-range repulsion --- the well-known Tonks-Girardeau (TG) limit --- the static structure factor never exceeds unity. In fact, $S_{\rm TG}(q)$ increases linearly from zero to unity, which is reached at $q=2k_F$, with $S_{\rm TG}(q)\equiv 1$ for all $q\geq 2k_F$. 

Quite remarkably, the line shift at high- momentum~[Eq.~\eqref{eq:shift}] in the low-density limit $na^3 \ll 1$ agrees with an old result by Beliaev~\cite{beliaev58b}, who presents a calculation of the boson Green's function, the poles of which coincide with the resonances of the dynamic structure factor in the symmetry-broken phase~\cite{griffin93}. Beliaev derives an expression for the single-particle energy in terms of the two-particle scattering amplitude $f({\bf q}/2,-{\bf q}/2)$. The high-momentum limit of his expression reads~\cite{beliaev58b}
\begin{align}
\hbar \omega_{\bf q} &\stackrel{q\to \infty}{\to}  \varepsilon_{\bf q} + {\rm Re} \Bigl[2 f\Bigl(\frac{\bf q}{2}, -\frac{\bf q}{2}\Bigr) - f({\bf 0}, {\bf 0})\Bigr] .
\end{align}
Using the expression for the two-body scattering amplitude, $f({\bf q}, -{\bf q}) = 4 \pi \hbar^2 n m^{-1}/(a^{-1} + i q)$, this is in agreement with our result in Eq.~\eqref{eq:shift} in the weak interaction limit, where $\mathcal{C}_2 = (4 \pi n a)^2$. Our work thus generalizes Beliaev's result to arbitrary scattering lengths. In fact, our result is universal in that it does not depend on temperature or even the thermodynamic phase of the gas. Quite generally, Eq.~\eqref{eq:shift} separates a functional dependence on momentum, which is essentially determined by few-body physics, from the probability density to find two bosons in close proximity, which is parametrized by the contact parameter $\mathcal{C}_2$.

\section{Dynamic structure factor of Bose gases beyond Bogoliubov}\label{sec:diagrammatics}

The aim of this section is to construct a simple many-body theory for the dynamic structure factor of an interacting Bose gas that is consistent with the exact high-momentum form provided by the combination of IA and OPE discussed in Sec.~\ref{sec:asymptotics}, and which provides an accurate description of the dynamic structure factor for all probe energies and wavelengths. It turns out that a simple one-loop approximation to the density response is not accurate even for a weakly interacting Bose gas, and, in particular, does not capture the OPE scaling behavior, which is linked to the breakdown of Bogoliubov theory at high momenta (which we already encountered when discussing the line shift). We show that this shortcoming of the one-loop approximation is corrected by including a Maki-Thompson-type correction, which describes the repeated scattering of two bosons in the high-momentum limit.

\subsection{Density and current response}\label{sec:response}

%++++++++++++++++++++++++++++++++++++++++
\begin{figure}[t!]
\scalebox{0.55}{\includegraphics{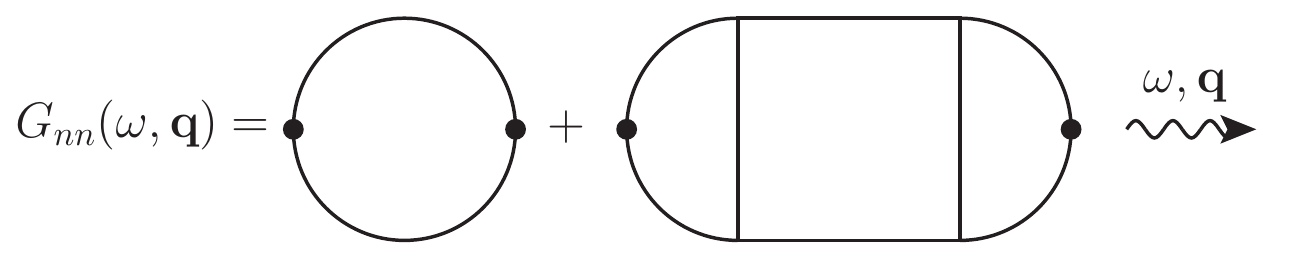}}
\caption{Many-body $T$ matrix approximation to the current response function. Continuous lines denote the Bogoliubov propagator, the square box the many-body $T$ matrix defined in Appendix~\ref{app:bogoliubov} and Fig.~\ref{fig:bethe}, and filled circles the density operator that inserts a frequency $\omega$ and a wave vector ${\bf q}$. The first diagram denotes the one-loop Bogoliubov approximation of Eqs.~\eqref{eq:cur1} and~\eqref{eq:cur2}. The second diagram denotes the $T$ matrix insertion and is given by Eq.~\eqref{eq:Tmatrixinsertion}.}
\label{fig:densityTmatrix}
\end{figure}
%++++++++++++++++++++++++++++++++++++++++

We choose the current response function as a starting point and define the density response in terms of the longitudinal current response function
\begin{align}
{\rm Im} \, \chi(\omega, {\bf q}) &= \frac{q^2}{\omega^2} {\rm Im} \, \chi_{jj}^L , \label{eq:ward}
\end{align}
where the current response is defined as
\begin{align}
&\chi_{ij}(i\omega_n,{\bf q}) = \frac{1}{\hbar V} \int_0^{\hbar\beta} d\tau \, e^{i\omega_n \tau} \langle T_\tau \hat{j}_{i \bf q}^{}(\tau) \hat{j}_{i \bf -q} \rangle \nonumber \\
&= \frac{q_iq_j}{q^2} \chi_{jj}^L(i\omega_n, {\bf q}) + \Bigl(\delta_{ij} - \frac{q_iq_j}{q^2}\Bigr) \chi_{jj}^T(i\omega_n, {\bf q}) ,
\end{align}
which we decompose in the second line into a scalar longitudinal and transverse part. The simplest approximation to the multiphonon part of the current response is the one-loop diagram in Fig.~\ref{fig:densityTmatrix}, which dates back to work by Fetter~\cite{fetter70} and Talbot and Griffin~\cite{talbot83}. In Fig.~\ref{fig:densityTmatrix}, the line denotes the Bogoliubov propagator defined as (for all details of the field theory in the condensed phase, see Appendix~\ref{app:bogoliubov})
\begin{align}
G(i\omega_n,{\bf q}) &= \begin{pmatrix} G_{11} &G_{21} \\ G_{12} &G_{22}\end{pmatrix} \nonumber \\
&= \frac{i\hbar\omega_n \sigma_3 + (\varepsilon_{\bf q} + gn_0) - g n_0 \sigma_1}{(i\hbar\omega_n)^2 - E_{\bf k}^2} ,\label{eq:propagator}
\end{align}
where $E_{\bf q} = \sqrt{\varepsilon_{\bf q} (\varepsilon_{\bf q} + 2 gn_0)}$ and $\sigma$ are the Pauli matrices. The one-loop diagram in Fig.~\ref{fig:densityTmatrix} contains two separate contributions that involve either the normal or the anomalous propagators,
\begin{align}
&\chi_{jj}^{L,a} \nonumber \\
&= \frac{1}{\beta V} \sum_{i\Omega_n, \bf k} F_{\bf k,q}^2 G_{11}(i\Omega_n + i\omega_n, {\bf k}+{\bf q}) G_{11}(i\Omega_n, {\bf k}) \nonumber \\
&\quad\quad + (\omega_n \to - \omega_n) \nonumber \\
&= \frac{1}{V} \sum_{\bf k} F_{\bf k,q}^2 \biggl\{\bigl[u_1^2 u_2^2 R_1(i\omega_n) + v_1^2 v_2^2 R_1(-i\omega_n)\bigr] \nonumber \\
&\qquad- \bigl[u_1^2 v_2^2 R_2(i\omega_n) + v_1^2 u_2^2 R_2(-i\omega_n)\bigr]\biggr\} + (\omega_n \to - \omega_n) \label{eq:cur1} ,
\end{align}
and
\begin{align}
&\chi_{jj}^{L,b} \nonumber \\
&= - \frac{1}{\beta V} \sum_{i\Omega_n, \bf k} F_{\bf k,q}^2 G_{21}(i\Omega_n + i\omega_n, {\bf k}+{\bf q}) G_{12}(i\Omega_n, {\bf k}) \nonumber \\
&\qquad + (i \omega_n \to - i \omega_n) \nonumber \\
&= - \frac{2}{V} \sum_{\bf k} F_{\bf k,q}^2 \biggl\{u_1 u_2 v_1 v_2 \bigl[ R_1(i\omega_n) + R_1(-i\omega_n) \nonumber \\
&\qquad- R_2(i\omega_n) - R_2(-i\omega_n)\bigr]\biggr\} \label{eq:cur2} ,
\end{align}
where we abbreviate $u_1 = u_{\bf k+q}$, $u_2 = u_{\bf k}$, $v_1 = v_{\bf k+q}$, and $v_2 = v_{\bf k}$ with the Bogoliubov coherence factors
\begin{align}
u_{\bf q}^2 &= \frac{\varepsilon_{\bf q} + gn_0}{2 E_{\bf q}} + \frac{1}{2} , \qquad v_{\bf q}^2 = \frac{\varepsilon_{\bf q} + gn_0}{2 E_{\bf q}} - \frac{1}{2} \label{eq:coherencefactor} ,
\end{align}
define the current matrix element $F_{\bf k,q} = \frac{\hbar}{2m} (2 {\bf k} \cdot \hat{\bf q} + q)$ ($\hat{\bf q}$ is the unit vector in the direction of ${\bf q}$), and introduce
\begin{align}
R_1(i\omega_n) &= \frac{f(E_{\bf k+q}) - f(E_{\bf k})}{i \hbar \omega_n - (E_{\bf k+q} - E_{\bf k})} \label{eq:R1} \\
R_2(i\omega_n) &= \frac{1 + f(E_{\bf k+q}) + f(E_{\bf k})}{i \hbar \omega_n - (E_{\bf k+q} + E_{\bf k})} \label{eq:R2}  .
\end{align}
At zero temperature, only the second term $R_2$ is nonzero.

%++++++++++++++++++++++++++++++++++++++++
\begin{figure}[t!]
\subfigure[]{\scalebox{0.4}{\raisebox{0.5cm}{\includegraphics{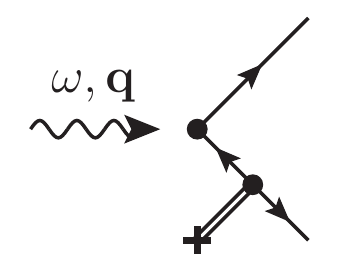}}}}\qquad
\subfigure[]{\scalebox{0.4}{\includegraphics{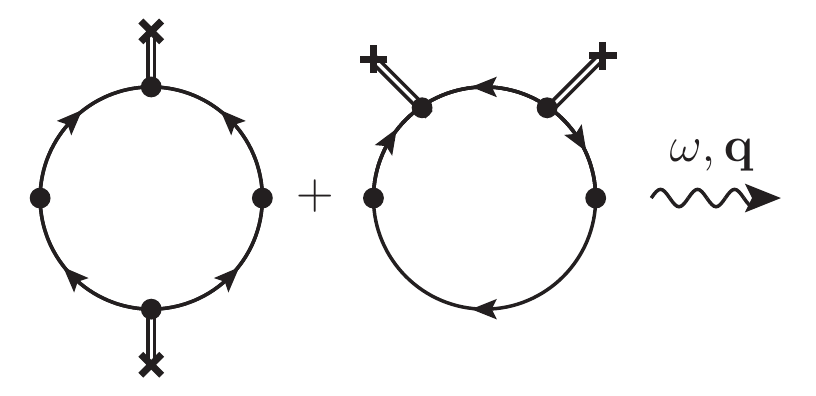}}}\quad
\subfigure[]{\scalebox{0.4}{\raisebox{0.75cm}{\includegraphics{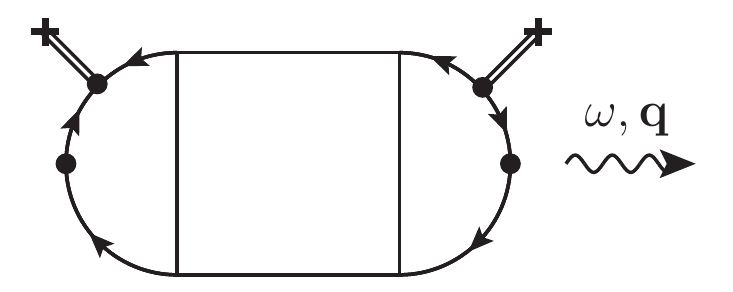}}}}
\caption{(a) Elementary process where two particles with large and opposite momentum are expelled from the condensate and couple to the density probe. The continuous lines are free-particle propagators. (b) Diagrams corresponding to the free-particle high-momentum limit of the one-loop Bogoliubov diagram in Fig.~\ref{fig:densityTmatrix}. (c) Diagram that describes the high-momentum limit of the diagram with $T$ matrix insertion in Fig.~\ref{fig:densityTmatrix}.}
\label{fig:OPE}
\end{figure}
%++++++++++++++++++++++++++++++++++++++++

Let us discuss the response function in the limit of large momentum transfer $\hbar {\bf q}$, where the scattered atom behaves as a free particle; i.e., $E_{\bf k+q} \approx \varepsilon_{\bf k+q}$. In line with the discussion in Sec.~\ref{sec:asymptotics} and Fig.~\ref{fig:regimes}, there are two distinct regimes depending on the initial-state atom: first, its energy can be much smaller than the final state energy, $E_{\bf k} \ll \varepsilon_{\bf k+q}$, and second, it can be of comparable magnitude; i.e., the Bragg pulse scatters off a pair of atoms with large momentum. In the first (quasifree) case, we neglect the contribution of Eq.~\eqref{eq:cur2} to the current response as a free particle does not have an off-diagonal propagator component. Equation~\eqref{eq:cur1} simplifies as follows: we set $u_1 = 1$ and $v_1 = 0$ and expand $F({\bf k}, {\bf q}) = \hbar q/2m$. In Eqs.~\eqref{eq:R1} and~\eqref{eq:R2}, we neglect the contribution $f(\varepsilon_{\bf k+q})$, and the denominator in both equations is approximately $i \hbar \omega_n - (\varepsilon_{\bf k+q} - \varepsilon_{\bf k})$. The remaining expression contains only the Bogoliubov expression for the incoherent part of the momentum distribution, $n({\bf k}) = v_{\bf k}^2 + (u_{\bf k}^2 + v_{\bf k}^2) f(\varepsilon_{\bf k})$. Substituting the results in Eq.~\eqref{eq:ward} and using $\hbar \omega \approx \varepsilon_{\bf q}$, we immediately arrive at the impulse approximation of Eq.~\eqref{eq:defIA} with the Bogoliubov momentum distribution. The dynamic structure factor at zero temperature can then be computed analytically, and we obtain for the scaling function $J_{\rm IA}$ defined in Eq.~\eqref{eq:IA}
\begin{align}
J_{\rm IA}^{\rm (Bog)}(Y) = \frac{1}{16\pi^2} (1 + Y^2 - |Y| \sqrt{2+Y^2})\label{eq:bogoliubovJIA}
\end{align}
with the scaling variable $Y = m \xi (\hbar \omega - \varepsilon_{\bf q})/\hbar^2 q$. The large-$Y$ tail agrees with the exact result of Eq.~\eqref{eq:JIAasym}, where here $\mathcal{C}_2 = 1/(4\xi^4)$. Likewise, the small-$|Y|$ cusp agrees with the exact result~[Eq.~\eqref{eq:JIAsmallY}], where we use $\xi = \hbar/\sqrt{2} m c_s$ in Bogoliubov theory.
 
In the second limit in which both initial and final states behave as a free particle, $E_{\bf k} \approx \varepsilon_{\bf k}$ and $E_{\bf k+q} \approx \varepsilon_{\bf k+q}$ (the multiparticle regime), the leading-order asymptotic is obtained by expanding the coherence factors in Eqs.~\eqref{eq:cur1} and~\eqref{eq:cur2} as $u_{\bf p} = 1$ and $v_{\bf p} = gn_0/2\varepsilon_{\bf p}$. Terms involving $f(\varepsilon_{\bf p})$ are exponentially suppressed and can be dropped. Without writing the explicit result, we note that this response corresponds to a process where two particles with large and opposite momentum are emitted from the condensate, one of which couples to the Bragg beam. This process is shown in Fig.~\ref{fig:OPE}(a), with Fig.~\ref{fig:OPE}(b) showing the contributions to the current response, where the lines are propagators of free particles with parabolic dispersion. The coupling to the condensate arises as a self-energy correction to the free-particle lines.

%++++++++++++++++++++++++++++++++++++++++
\begin{figure}[t]
\scalebox{0.7}{\includegraphics{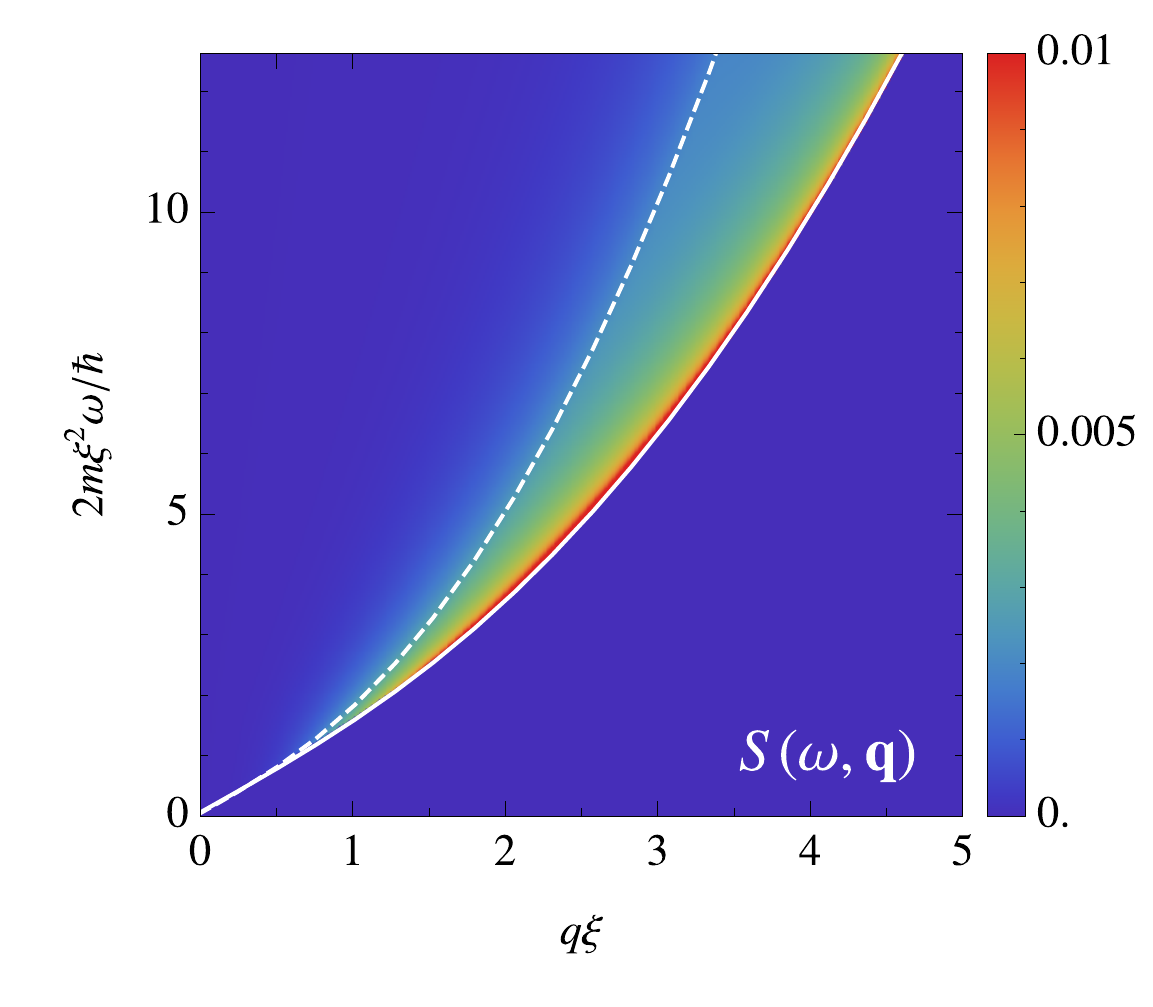}}
\caption{Density plot of the dynamic structure factor as predicted by the $T$ matrix approximation of Sec.~\ref{sec:diagrammatics}. The response starts above a two-phonon threshold $\hbar \omega = 2 E_{{\bf q}/2}$ indicated by the white continuous line. The position of the coherent Bogoliubov peak at $\hbar \omega = E_{\bf q}$ is shown by a white dashed line.}
\label{fig:structurefactor}
\end{figure}
%++++++++++++++++++++++++++++++++++++++++

%++++++++++++++++++++++++++++++++++++++++
\begin{figure*}[t]
\scalebox{0.6}{\includegraphics{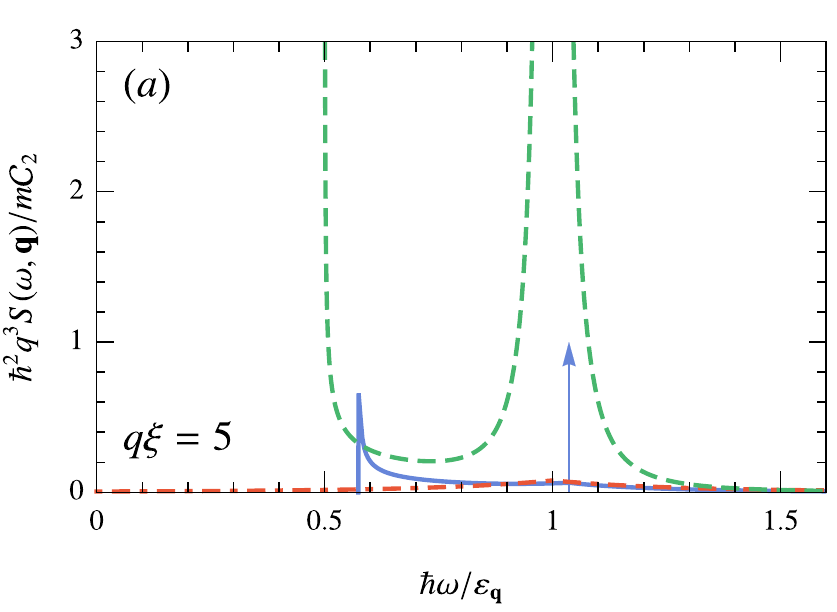}}\hspace{0.3cm}
\scalebox{0.6}{\includegraphics{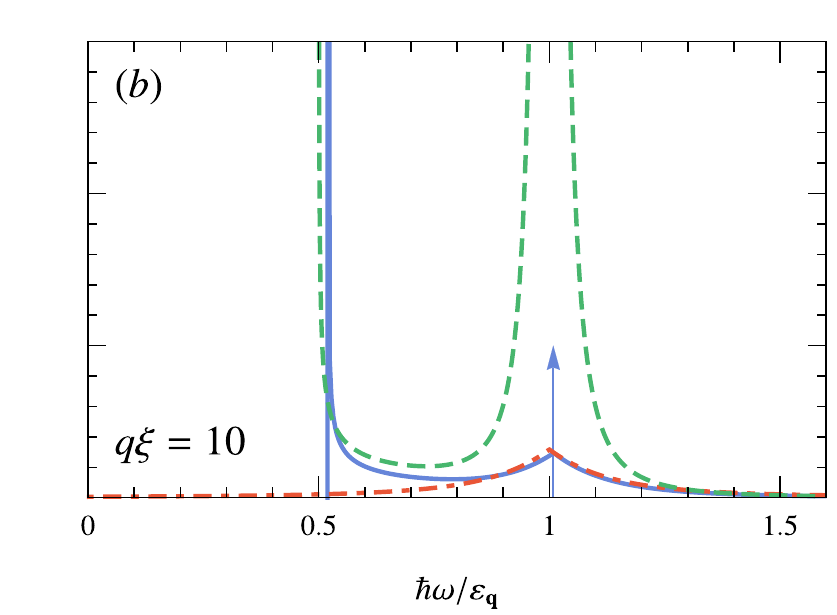}}\hspace{0.3cm}
\scalebox{0.6}{\includegraphics{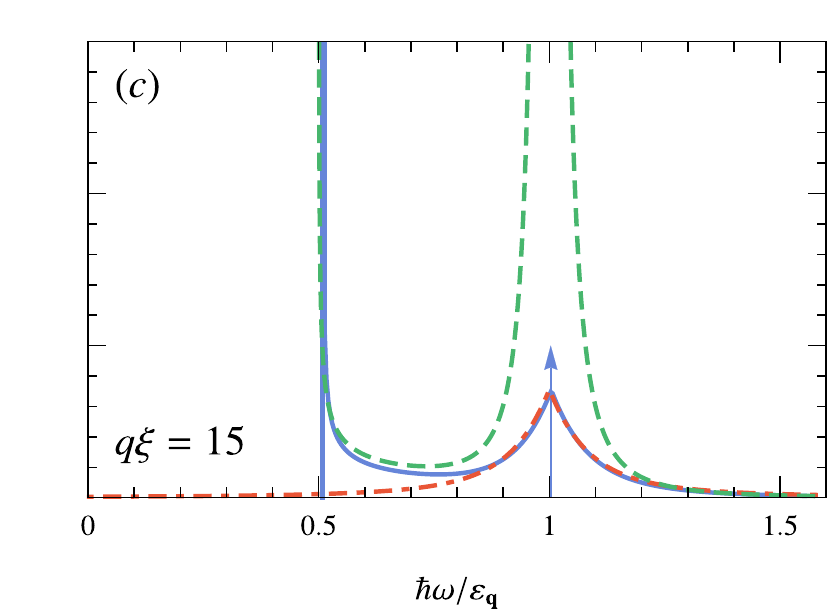}}
\scalebox{0.6}{\includegraphics{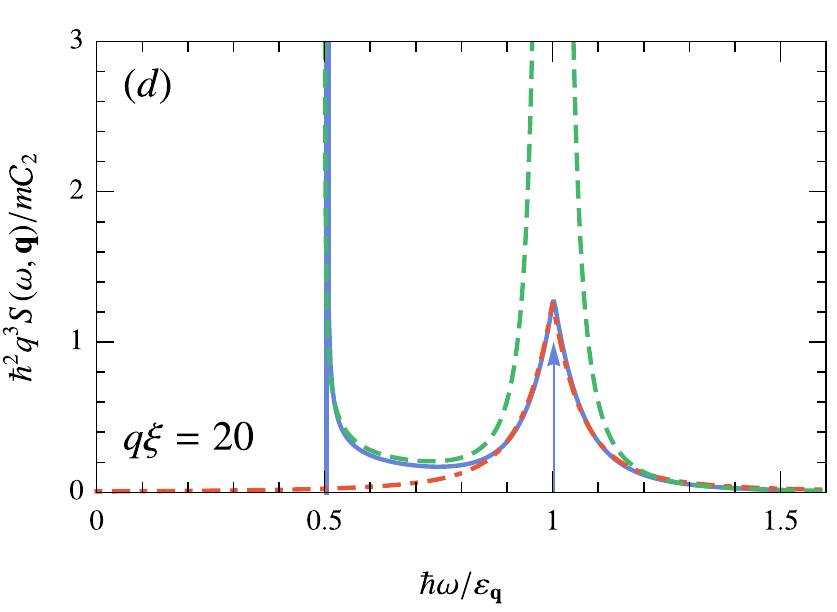}}\hspace{0.3cm}
\scalebox{0.6}{\includegraphics{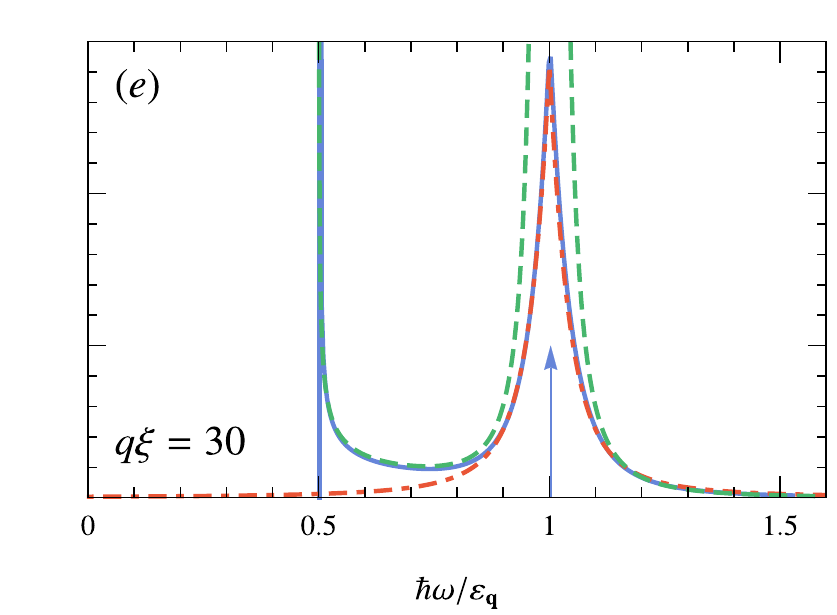}}\hspace{0.3cm}
\scalebox{0.6}{\includegraphics{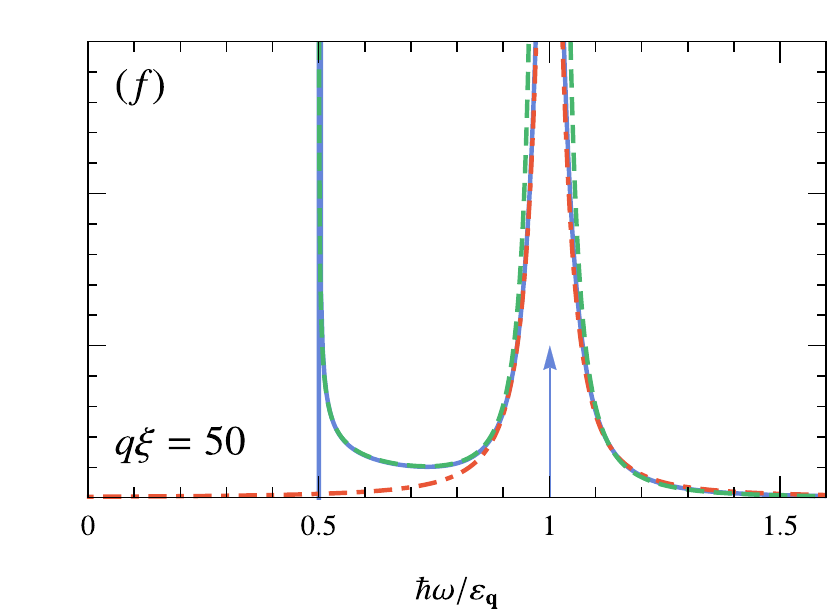}}
\caption{Dynamic structure factor as a function of frequency for increasing momentum $q\xi=5,10,15,20,30$, and $50$ [$(a)$-$(f)$] (continuous blue lines). The plots clearly show the crossover to the universal scaling form predicted by the OPE (dashed green lines) at high momentum, shown in full in Fig.~\ref{fig:JOPE}. Near the one-particle peak, the dynamic structure factor is described accurately by the impulse approximation (red dot-dashed line). The vertical arrows (not to scale) indicate the position of the condensate delta peak on top of the incoherent background, which here is centered at the Bogoliubov energy $E_{\bf q} = \sqrt{\varepsilon_{\bf q} (\varepsilon_{\bf q} + 2gn)}$.}
\label{fig:crossover}
\end{figure*}
%++++++++++++++++++++++++++++++++++++++++

It is immediately clear that this one-loop description of the free-particle limit must be incomplete, because it neglects the interaction between the initial and the final states. They are accounted for by including the many-body $T$ matrix in the current response, as depicted by the second diagram in Fig.~\ref{fig:densityTmatrix}. A definition and explicit expression for the $T$ matrix is given in Appendix~\ref{app:bogoliubov}. Formally, the correction corresponds to a Maki-Thompson correction on the Bogoliubov level. The explicit form $\chi_{jj}^{L,{\rm MT}}$ can be written as a matrix bilinear:
\begin{align}
\chi_{jj}^{L,{\rm MT}} &= {\bf f}^T T {\bf f} , \label{eq:Tmatrixinsertion}
\end{align}
where the $T$ matrix is defined in Appendix~\ref{app:bogoliubov} and
\begin{align}
{\bf f} &= \frac{2}{\beta V} \sum_{i\Omega_n, \bf k} F_{\bf k,q}
\begin{pmatrix}
G_{12}(i\Omega_n + i\omega_n, {\bf k}+{\bf q}) G_{22}(i\Omega_n, {\bf k}) \\[1ex]
G_{22}(i\Omega_n + i\omega_n, {\bf k}+{\bf q}) G_{21}(i\Omega_n, {\bf k})
\end{pmatrix} . \label{eq:f}
\end{align}
Here, we define
\begin{align}
&f_1(i\omega_n, {\bf q}) \nonumber \\
&\quad = \frac{2}{V} \sum_{\bf k} F_{\bf k,q} \biggl\{- u_1 v_1 \bigl[v_2^2 R_1(i\omega_n) + u_2^2 R_1(-i\omega_n)\bigr] \nonumber \\
&\quad \quad + u_1 v_1 \bigl[u_2^2 R_2(i\omega_n) + v_2^2 R_2(-i\omega_n)\bigr]\biggr\} \label{eq:cur3} .
\end{align}
while $f_1(i \omega_n, {\bf q}) = - f_2^*(i \omega_n, {\bf q})$. Using the explicit form of the $T$ matrix, we obtain two additional contributions to the current response function:
\begin{align}
\chi_{jj}^{\rm L,MT,a} &= \frac{\frac{1}{8 \pi \hbar^2 a} - \Pi_{d}^*}{(\frac{m}{8 \pi \hbar^2 a} - \Pi_{d}^*) (\frac{m}{8 \pi \hbar^2 a} - \Pi_{d}^{}) - \Pi_{c}^2} f_1(i\omega_n)^2 \nonumber \\
&\quad+ (i\omega_n \to - i\omega_n)
\end{align}
and
\begin{align}
\chi_{jj}^{\rm L,MT,b} &= - \frac{\Pi_{c}}{(\frac{m}{8 \pi \hbar^2 a} - \Pi_{d}^*) (\frac{m}{8 \pi \hbar^2 a} - \Pi_{d}^{}) - \Pi_{c}^2} \nonumber \\
&\quad \times f_1(i\omega_n) f_1(-i\omega_n) + (i\omega_n \to - i\omega_n) ,
\end{align}
where
\begin{align}
\Pi_{c}(i\omega_n, {\bf q}) &= \frac{1}{V} \sum_{\bf k} \biggl\{u_1 u_2 v_1 v_2 \bigl[ R_1(i\omega_n) + R_1(-i\omega_n) \nonumber \\
&\quad- R_2(i\omega_n) - R_2(-i\omega_n)\bigr]\biggr\} \label{eq:Pic}
\end{align}
and
\begin{align}
&\Pi_{d}(i\omega_n, {\bf q}) = \frac{1}{V} \sum_{\bf k} \biggl\{\bigl[u_1^2 v_2^2 R_1(i\omega_n) + v_1^2 u_2^2 R_1(-i\omega_n)\bigr] \nonumber \\
&\quad- \bigl[u_1^2 u_2^2 R_2(i\omega_n) + v_1^2 v_2^2 R_2(-i\omega_n)\bigr] -\frac{1}{2\varepsilon_{\bf k}}\biggr\} \label{eq:Pid} .
\end{align} 
To return to the high-momentum limit, we see that the $T$ matrix correction does not contribute to the IA result since ${\bf f}$ in Eq.~\eqref{eq:f} involves an off-diagonal propagator. The free-particle limit, however, receives a correction which corresponds to the diagram shown in Fig.~\ref{fig:OPE}(c). 

Before concluding this section, note that we do not take into account self-energy corrections to the Bogoliubov propagators, which would require a renormalization of the chemical potential, as discussed in Refs.~\cite{braaten97,braaten99}. As discussed in Ref.~\cite{hofmann11}, such self-energy corrections contribute to the asymptotic response as well (they correspond to processes where two particles are emitted from the condensate and scatter before they couple to the probe). These corrections, however, are subleading in $q$ [i.e., they are suppressed as $\mathcal{O}(1/qa)$]~\cite{hofmann11} and provide corrections to the asymptotic result in Eq.~\eqref{eq:opedynstruc}. We do not include them here.

\subsection{Results}\label{sec:results}

%++++++++++++++++++++++++++++++++++++++++
\begin{figure*}[t]
\subfigure[]{
\scalebox{0.62}{\includegraphics{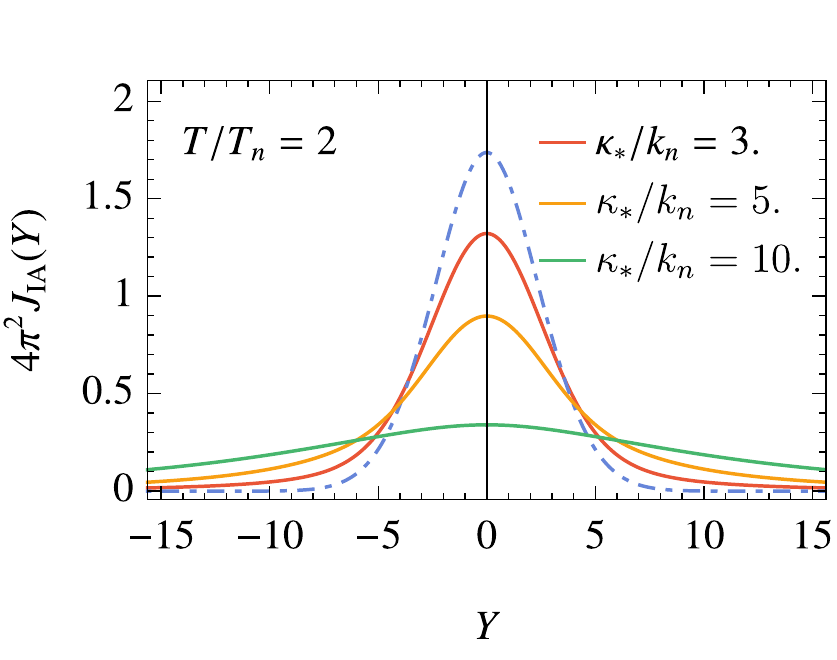}}}\qquad
\subfigure[]{
\scalebox{0.62}{\includegraphics{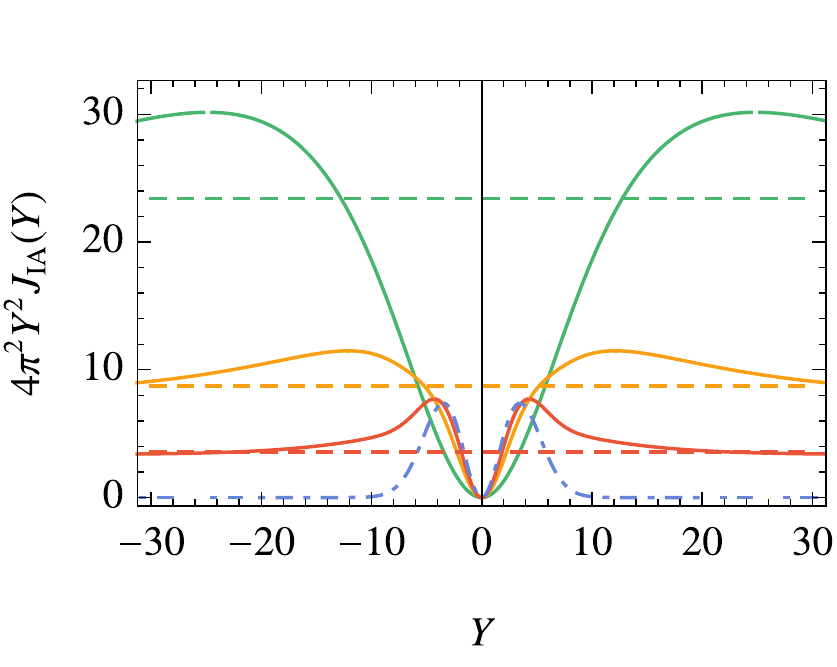}}}\qquad
\subfigure[]{
\scalebox{0.63}{\includegraphics{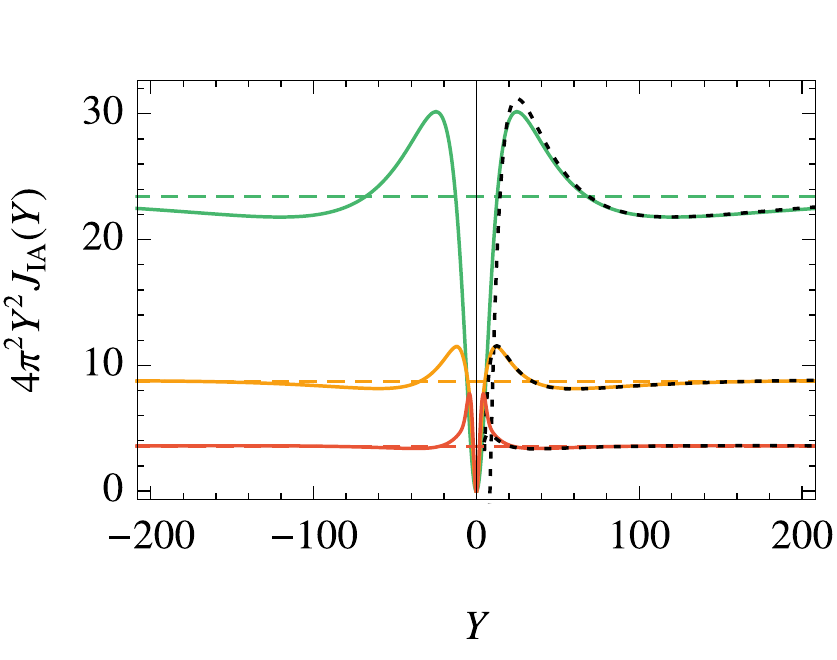}}}
\caption{(a) Impulse approximation $Y$-scaling function $J_{\rm IA}$ for the nondegenerate Bose gas at unitarity and $T/T_n=2$. The continuous red, orange, and green lines show the scaling function for three different values of the three-body parameter $\kappa_*/k_n=3,5$, and $10$. For comparison, we include the impulse approximation for the noninteracting gas at the same temperature. (b) The impulse approximation shows heavy tails for large $Y$ which are in good agreement with the leading-order two-body Tan relation and a subleading three-body correction. (c) The three-body tail is indicated by black dotted lines.}
\label{fig:YscalingBose}
\end{figure*}
%++++++++++++++++++++++++++++++++++++++++

In the following, we present the results of our $T$ matrix approximation to the dynamic structure factor, restricting our attention to the zero-temperature case and $a\gg \xi$ for simplicity, where the coherence length $\xi = \hbar/\sqrt{2mgn}$ sets the unit of wave number. We perform the integrations in Eqs.~\eqref{eq:cur1},~\eqref{eq:cur2},~\eqref{eq:cur3},~\eqref{eq:Pic}, and~\eqref{eq:Pid} numerically using the \textsc{cuba} library~\cite{hahn05}. These one-loop results are then used to obtain the dynamic structure factor as described in Sec.~\ref{sec:response}. Figure~\ref{fig:structurefactor}  shows the result for the dynamic structure factor as a function of wave number and frequency. The position of the single-particle peak lies at $\hbar \omega = E_{\bf q}$ shown as a white dashed line.  As is apparent from the figure, the threshold for multiphonon excitations lies at $\hbar \omega = 2 E_{{\bf q}/2}$, indicated by a white solid line. The incoherent spectral weight is due to excitations of two and more particles. For a density probe with energy $\hbar \omega$ and wave vector ${\bf q}$, the minimum threshold energy to create two excitations with dispersion $E_{\bf q}$ (where $E_{\bf q}$ is a convex function of momentum) is $\hbar \omega = 2 E_{{\bf q}/2}$. At small wave vectors ${\bf q}$, this threshold energy coincides with the Bogoliubov mode, whereas at large wave vector, it lies below the single-particle energy at $\hbar \omega = \varepsilon_{\bf q}/2$.  The incoherent spectral weight is strongly concentrated near the two-particle threshold, even for momenta larger than $\xi^{-1}$ where the single-particle spectrum deviates from the linear-in-$q$ Bogoliubov form. Consistent with the general arguments by Feynman~\cite{feynman54} as well as Miller, Pines, and Nozi\`{e}res~\cite{miller62}, the incoherent spectral weight decreases rapidly at small wave vector. At high momentum, the integrated weight gives the density of the noncondensed atoms~\cite{hohenberg66}.

Figure~\ref{fig:crossover} shows the dynamic structure factor as a function of $\hbar \omega /\varepsilon_{\bf q}$ for six different momenta $q\xi = 5,10,15,20,30$, and $50$ [Figs.~\ref{fig:crossover}(a)--(f), respectively] as blue continuous lines. The OPE scaling form of Eq.~\eqref{eq:opedynstruc} and Fig.~\ref{fig:JOPE} is shown in Fig.~\ref{fig:crossover} as a green dashed line. From Fig.~\ref{fig:crossover}, it can be seen very clearly that at high momentum the dynamic structure factor converges to the OPE result. The full numerical solution is seen to match the scaling prediction very accurately away from the single-particle peak, and the region near $\hbar \varepsilon_{\bf q}$ where the theories differ shrinks with increasing momentum. We also compare the results of Fig.~\ref{fig:crossover} to the IA computed in Eq.~\eqref{eq:bogoliubovJIA} (red dot-dashed line). The IA describes the dynamic structure factor near the single-particle peak very accurately. As the momentum in increased, it is apparent from Fig.~\ref{fig:crossover} that the IA scaling crosses over to the OPE scaling form. This calculation illustrates how the asymptotic scaling emerges in the Bose gas.

\section{$Y$ scaling of the nondegenerate Bose gas}\label{sec:Ythermal}

The IA scaling function $J_{\rm IA}(Y)$, Eq.~\eqref{eq:JIA}, depends only on the momentum distribution. On the one hand, this opens the possibility to probe the momentum distribution through measurements of the dynamic structure factor. On the other hand, calculations of the momentum distribution can be used to compute the scaling form of the dynamic structure factor [as was done in Eq.~\eqref{eq:bogoliubovJIA} on the Bogoliubov level]. In this section, we compute the scaling function in the nondegenerate regime, for which accurate calculations of the momentum distribution were performed in Ref.~\cite{barth15}.

In the nondegenerate limit, where the interparticle spacing is much larger than the thermal wavelength $\lambda_T = \hbar \sqrt{2\pi/mT}$, i.e., $n\lambda_T^3 \ll 1$, the fugacity $z=e^{\beta \mu}$ is small, thus providing a systematic expansion parameter. Reference~\cite{barth15} computes the momentum distribution up to third order in $z$:
\begin{align}
n(k) &= z n_1(k) + z^2 n_2(k) + z^3 n_3(k) + \mathcal{O}(z^4) .
\end{align}
The first term, $n_1(k)$, is the Boltzmann distribution of a noninteracting gas. The second and third terms, $n_2(k)$ and $n_3(k)$, respectively, take into account two-body correlations and three-body correlations exactly. Three-body correlations are manifest through the dependence on the three-body parameter $\kappa_*$, a wave vector that sets the energy of the lowest Efimov trimer $E_T=\hbar^2\kappa_*^2/2m$~\cite{braaten06} and that depends in an approximately universal way on the van der Waals length $\ell_{\rm vdW}$ as $\kappa_* \sim 0.2/\ell_{\rm vdW}$~\cite{wang12,schmidt12}. Typical values for $^{85}$Rb are, for example, $\kappa_* = 30 \mu {\rm m}^{-1}$~\cite{wild12}, $l_{\rm vdW} = 160 a_0$~\cite{vankempen02}, and $n = 5 \times 10^{12} {\rm cm}^{-1}$~\cite{makotyn14}, implying $k_n \ell_{\rm vdW} \sim 10^{-2}$ (as is expected for a dilute system~\cite{zwerger16}) and $\kappa_*/k_n \sim 5-10$. We use the results for the momentum distribution presented in Ref.~\cite{barth15} in Eq.~\eqref{eq:JIA} to compute the scaling function $J_{\rm IA}(Y)$ as a function of the scaling parameter $Y = m \lambda_T (\hbar \omega - \varepsilon_{\bf q})/\hbar^2 q$ (note that the characteristic length scale is set by the thermal wavelength $\lambda_T$ in the nondegenerate gas). The results of this calculation are shown in Fig.~\ref{fig:YscalingBose} at fixed temperature $T/T_n=2$ for three values of the three-body parameter $\kappa_*/k_n = 3,5$, and $10$ (continuous lines), where $k_n=(6\pi^2n)^{1/3}$ and $T_n=\hbar^2k_n^2/2m$. For comparison, we include the noninteracting result as a dash-dotted line. Figure~\ref{fig:YscalingBose}(b) shows the scaling function multiplied by $Y^2$ to extract the power-law tail of the distribution, Eq.~\eqref{eq:JIAasym}. The magnitude of the tail given by $\mathcal{C}_2/2$ is shown by dashed lines with a numerically computed two-body contact parameter~\cite{barth15}. It turns out that there exists an analytical expression for the subleading scaling of a Bose gas as well: using the subleading asymptotic behavior of the momentum distribution~[Eq.~\eqref{eq:tailC3}], we obtain a large-$Y$ tail of
\begin{align}
J_{\rm IA}(Y) &= \frac{\lambda_T^4 \mathcal{C}_2}{8 \pi^2 Y^2} \nonumber \\
&+ \frac{\lambda_T^5 A \mathcal{C}_3}{4 \pi^2 (9 + 4 s_0^2) |Y|^3} \biggl[2 s_0 \cos\Bigl(2 \phi + 2 s_0 \ln \frac{|Y|}{\kappa_* \lambda_T}\Bigr) \nonumber \\
&\qquad \qquad+ 3 \sin\Bigl(2 \phi + 2 s_0 \ln \frac{|Y|}{\kappa_* \lambda_T}\Bigr)\biggr] .
\end{align}
This subleading correction to the tail is apparent in Fig.~\ref{fig:YscalingBose}(c), and we include it as a black dotted line. It is noticeable that the onset of the three-body correction lies already at moderate $Y$-parameter values. A measurement of the scaling function could thus be a reliable way to detect three-body physics.

\section{Application to Fermi gases}\label{sec:fermi}

Much of the discussion of the previous sections carries over directly to strongly interacting two-component Fermi gases with minor modifications. This is particularly interesting as the dynamic structure factor of strongly interacting Fermi gases has been measured in several experiments~\cite{veeravalli08}, and, in particular, the high-momentum structure has been measured in some detail~\cite{hoinka12,hoinka13}. This section lists the changes that arise when extending our results to the Fermi gas case. The form of the OPE near the single-particle peak is
\begin{align}
\lim_{Z\to 0} J_{\rm OPE}(Z) &= \frac{1}{\pi^2 Z^2} \biggl[1 + \frac{2}{1+(qa/2)^2}\biggr]  + \cdots .
\end{align}
While the coefficient is different by a factor of $2$ from the Bose gas result in Eq.~\eqref{eq:limitOPEBose}, the general picture remains unchanged: the IA describes small deviations from the single-particle peak, and the OPE describes large deviations. It would be interesting if the crossover between IA and OPE could be detected by measuring the dynamic structure factor as in Ref.~\cite{hoinka12,hoinka13} for several large momenta.

\begin{figure}[t]
\subfigure[]{\scalebox{0.62}{\includegraphics{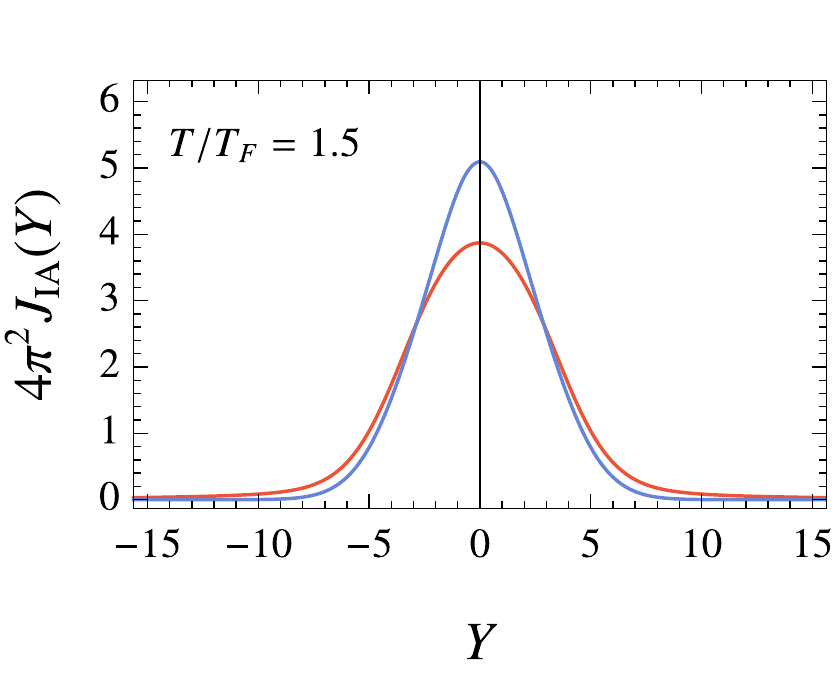}}\qquad}\qquad
\subfigure[]{\scalebox{0.63}{\includegraphics{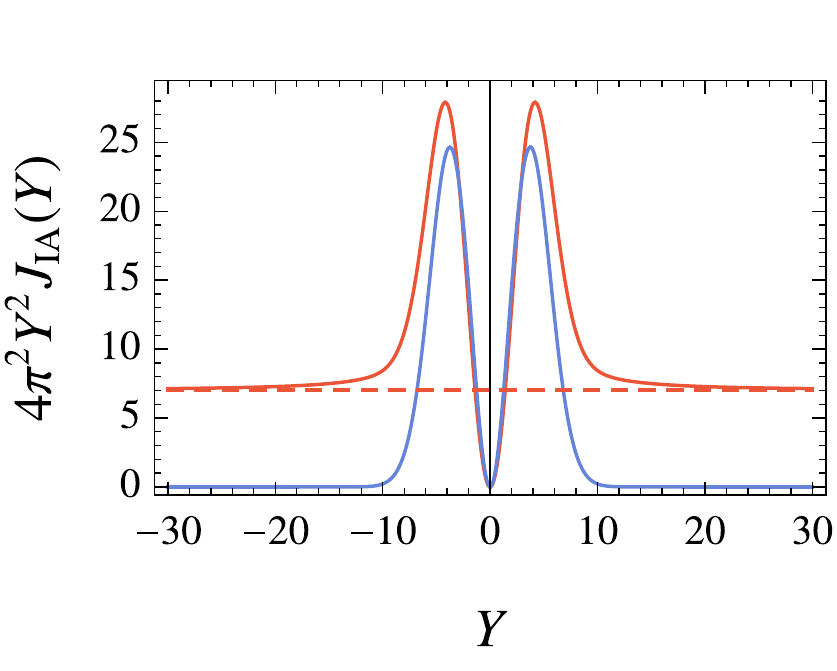}}\qquad}
\caption{(a) Impulse approximation $Y$-scaling function $J_{\rm IA}$ for the nondegenerate Fermi gas at unitarity. The red line denotes the scaling function of the interacting gas with $T/T_F=1.5$. For comparison, we include the scaling function of a noninteracting Fermi gas with the same density. (b) The power-law $1/Y^2$ tail Eq.~\eqref{eq:JIAasym} of the interacting gas. The dashed red line marks the asymptotic behavior with $\lambda_T^4 \mathcal{C}_2 = 14.045$.}
\label{fig:YscalingFermi}
\end{figure}

As for the partial sum rules, the IA results are unchanged, where we have to keep in mind that the density now refers to the total density of both spin components, for which the high-momentum tail is $n(k) = 2 \mathcal{C}_2/k^4$. The OPE contribution is
\begin{gather}
m_{-1}^{({\rm OPE})} = \frac{\pi \mathcal{C}_2}{4 \varepsilon_{\bf q} q} + \frac{\hbar^2 q}{2 \pi^2 m} \frac{\mathcal{C}_2}{\varepsilon_{\bf q} \eta} \\
m_{0}^{({\rm OPE})} = \frac{\mathcal{C}_2}{4 q} + \frac{\hbar^2 q}{2 \pi^2 m} \frac{\mathcal{C}_2}{\eta} \\
m_{1}^{({\rm OPE})} = \frac{\hbar^2 q}{2 \pi^2 m} \frac{\varepsilon_{\bf q} \mathcal{C}_2}{\eta} \\
m_{2}^{({\rm OPE})} = \frac{\hbar^2 q}{2 \pi^2 m} \frac{\varepsilon_{\bf q}^2 \mathcal{C}_2}{\eta} .
\end{gather}
The result for the line shift reads
\begin{align}
\hbar \omega &= \varepsilon_{\bf q} + \frac{\hbar^2 \mathcal{C}_2}{2 \pi m a n} \, \biggl[\frac{2}{1+ (qa/2)^2} - 1\biggr] .
\end{align}
Finally, we use the results of Ref.~\cite{barth15} to present the $Y$-scaling function in the impulse approximation at high temperature, which is shown in Fig.~\ref{fig:YscalingFermi}. Results for the interacting Fermi gas at temperature $T/T_F = 1.5$ are shown by a red continuous line. For comparison, we include the result for the noninteracting gas at the same temperature as a blue dashed line. The $Y^2$ tail is directly evident in Fig.~\ref{fig:YscalingFermi}(b). A log-periodic correction as for the Bose gas is not present as the mass-balanced Fermi gas does not show the Efimov effect.

%++++++++++++++++++++++++++++++++++++++++
%++++++++++++++++++++++++++++++++++++++++

\section{Summary and Conclusions}\label{sec:summary}

In summary, we discuss the dynamic structure factor of a strongly interacting quantum gas in the deep inelastic regime of large momentum transfer. As the main result of our work, we establish that the high-momentum structure near the single-particle peak is governed by two separate scaling regimes: small-energy deviations from the single-particle peak [where $\hbar \omega - \varepsilon_{\bf q} \sim \mathcal{O}(q)$] are described by the impulse approximation, whereas large-energy deviations [where $\hbar \omega - \varepsilon_{\bf q} \sim \mathcal{O}(q^2)$] are described by the operator product expansion. This provides a complete description of the high-momentum structure factor at all frequencies. As an important consistency check, we compute the four highest moment sum rules as well as a Borel sum rule directly from the asymptotic form and provide restricted sum rules that apply to the IA and the OPE regime. Furthermore, we employ the OPE to compute the interaction correction to the position of the single-particle peak in the dynamic structure factor in the high-momentum limit, which naturally indicates a backbending from positive to negative values as the scattering length is increased. These exact results, which depend on only the two-body contact as a nonperturbative parameter, provide a qualitative explanation of the experiments performed at JILA in 2008~\cite{papp08}. Moreover, they also pave the way for a more detailed analysis of ongoing measurements of Bragg spectra at high momenta for strongly interacting Bose gases in box potentials.

In the second part of the paper, we give a simple $T$ matrix approximation to the dynamic structure factor. This approximation is exact for a weakly interacting Bose gas, but we expect it to be accurate even for large scattering lengths. In particular, our calculation illustrates how the structure factor probes the unitary limit for large momentum (where $qa\gg 1$), even if the gas itself is weakly interacting (i.e., $na^3\ll 1$). This regime cannot be captured by a simple one-loop calculation even for the weakly interacting gas, but requires a $T$ matrix correction in the form of a Maki-Thompson diagram. At low momentum, the single-particle peak in the dynamic structure factor coincides with the onset of the multiparticle incoherent spectral weight, while at high momentum, the single-particle peak lies in the continuum. Drawing on previous results for the momentum distribution of the Bose gas in the nondegenerate regime, we present results for the IA scaling function at unitarity, which shows a strong dependence on the three-body parameter and could thus be a sensitive probe of three-body physics. The results on the high-momentum scaling are not restricted to Bose gases and can be easily generalized to Fermi gases.

Viewed from a more general perspective, our work provides a solution of the long-standing problem of treating final-state interactions beyond the impulse approximation.  For the special case of ultracold gases with zero-range interactions, we show that the operator product expansion allows a systematic disentangling of two-body, three-body, and higher-order contributions to deal with multiparticle effects where a large momentum is transferred to an increasing number of particles.  From a technical point of view, our method is analogous to that used in high-energy physics, where QCD interaction effects beyond the parton model may be included by a short-distance expansion of the current-current correlators which determine the scattering cross sections. Methods developed in this context can thus be applied successfully in ultracold gases, at energy scales many orders of magnitude below that of high-energy physics.

\begin{acknowledgments}
It is a pleasure to thank Sandro Stringari for a number of discussions in the initial stage of this work. We are also grateful to Zoran Hadzibabic and Raphael Lopes for keeping us informed about ongoing measurements on Bragg scattering at high momentum on $^{39}$K in box potentials. This work is supported by Gonville and Caius College, Cambridge (J.H.). 
\end{acknowledgments}

\appendix

\section{Operator-product expansion}\label{app:ope}

Here, we summarize some details of the operator-product expansion. We restrict our attention to terms in the density response OPE that contribute to the singularity near the single-particle resonance, the results of which are quoted in Eqs.~\eqref{eq:Jn}, \eqref{eq:limitOPEBose}, and~\eqref{eq:Pionshell}. A detailed discussion of the density response OPE can be found in Refs.~\cite{son10,hofmann11,goldberger12, nishida12}.

The leading-order OPE terms are formed of operators with lowest scaling dimension. The only important operators here are the boson creation and annihilation operators $\hat{\phi}^\dagger$ and $\hat{\phi}$ with scaling dimension $\Delta_\phi = 3/2$, the density operator with $\Delta_n = 3$, the operator $\hat{O}_t = \hat{\phi}^\dagger (i\hbar\partial_t + \frac{\hbar^2 \nabla^2}{2m}) \hat{\phi}$ with $\Delta_t = 5$, and the two-body contact operator $\hat{O}_c$, which in explicit form reads for a zero-range interaction with strength $g$
\begin{align}
{\hat O}_c &= \biggl(\frac{m g}{\hbar^2}\biggr)^2 \hat{\phi}^{2\dagger} \hat{\phi}^2 ,
\end{align}
where $g$ is related to the scattering length via
\begin{align}
\frac{1}{g} = \frac{m}{4 \pi \hbar^2 a} - \frac{1}{V} \sum_{\bf k} \frac{1}{2\varepsilon_{\bf k}} . \label{eq:ren}
\end{align}
The expectation value ${\cal C}_2 = \langle \hat{O}_c \rangle$ appears in the interaction energy density, which is ill-defined for zero-range interactions, giving rise to an anomalous dimension $\Delta_{\mathcal{C}_2}=4$.

Since the Wilson coefficients do not depend on the state, they are determined by computing the operator expectation values in Eq.~\eqref{eq:defOPE} between one- and two-particle states and by matching the result. For the one-particle state with energy $p_0$ and momentum ${\bf p}$, this reads pictographically as
\begin{align}
&\raisebox{-0.075cm}{\scalebox{0.6}{\includegraphics{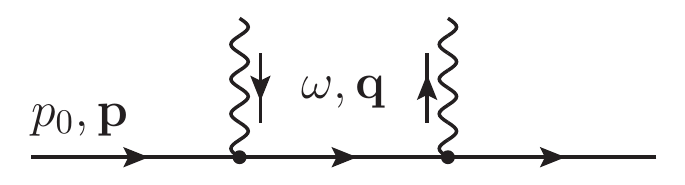}}} + (\omega,{\bf q} \to - \omega, - {\bf q}) \nonumber \\[2ex]
&\qquad= \sum_\ell W_\ell(\omega,{\bf q}) \times\raisebox{-0.075cm}{\scalebox{0.6}{\includegraphics{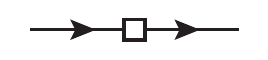}}} ,
\end{align}
where the vertex connected to the wavy line denotes the density insertion, the continuous line is the free-particle propagator at zero temperature $G_0(k_0, {\bf k}) = i/(k_0-\varepsilon_{\bf k} + i0)$, and the box is the insertion of a bilinear operator such as $\hat{O}_n$ or $\hat{O}_t$. The left-hand side is equal to $i G_0(p_0+\omega, {\bf p}+{\bf q}) + i G_0(p_0-\omega, {\bf p}-{\bf q})$. Expanding in $p_0$ and ${\bf p}$ and matching with one-body expectation values of bilinear operators, we obtain $W_n(\omega,{\bf q}) = i G_0(\omega, {\bf q}) + (\omega\to -\omega)$, cf. Eq.~\eqref{eq:Jn}, and $W_t(\omega, {\bf q}) = - G_0^2(\omega, {\bf q})$.

To obtain the Wilson coefficient of the contact operator, it is sufficient to match matrix elements for a two-particle state with zero external energy and momentum. The matrix element of the contact operator is $\langle 2p | \hat{O}_c | 2p \rangle = \bigl(\frac{m}{\hbar^2}\bigr)^2 {\cal A}$ with ${\cal A} = T_2(0,{\bf 0})$, where $T_2(\omega, {\bf q}) = - 8\pi\hbar^2/m(a^{-1} + i \sqrt{m\omega/\hbar - q^2/4})$. There are only two terms that contribute to the divergence near the single-particle peak. First, we have (the box denotes a $T$ matrix insertion $iT_2$)
\begin{align}
&W_c^{(1)}(\omega,{\bf q}) = \biggl(\frac{\hbar^2}{m}\biggr)^2 \frac{1}{{\cal A}^2} \times \raisebox{-0.35cm}{\scalebox{0.4}{\includegraphics{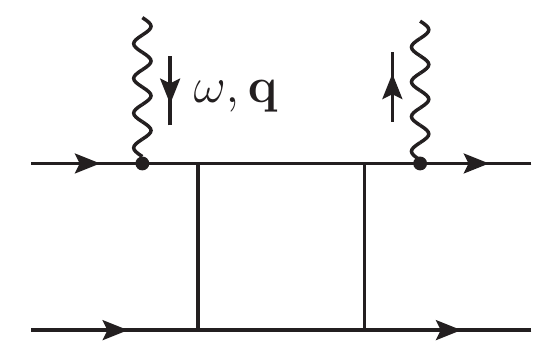}}} \nonumber \\
&\quad\stackrel{\hbar\omega\to\varepsilon_{\bf q}}{\to} - \frac{\hbar^2}{2\pi m a^2} \frac{1}{a^{-1} + i q/2} \frac{1}{(\hbar \omega - \varepsilon_{\bf q})^2}.
\end{align}
This term is singular because of the two Green's functions connecting the operator insertions and the scattering matrix and, can be interpreted as a vertex correction. In addition, there is a self-energy correction:
\begin{align}
W_c^{(2)}(\omega,{\bf q}) &= \biggl(\frac{\hbar^2}{m}\biggr)^2 \frac{1}{{\cal A}^2} \times \raisebox{-0.35cm}{\scalebox{0.4}{\includegraphics{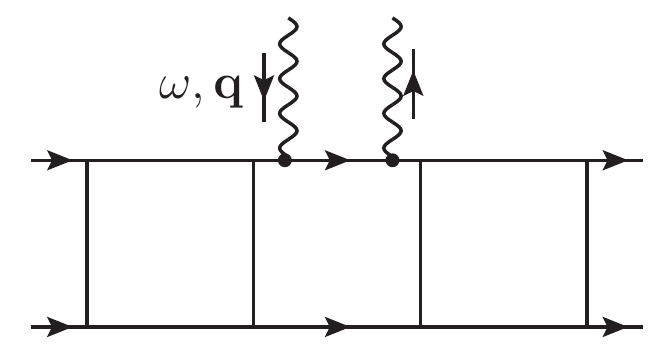}}} \nonumber \\
&\stackrel{\hbar\omega\to\varepsilon_{\bf q}}{\to} i \frac{\hbar^2 q}{8\pi m} \frac{1}{(\hbar \omega - \varepsilon_{\bf q})^2} .
\end{align}
There is a third contribution since we need to subtract the contribution of bilinear operators from the Wilson coefficient. This enters in two ways: first, through the expectation value of the bilinear insertion in the two-particle state (a divergent term), and, second, since the operator $\langle \hat{O}_t \rangle = \bigl(\frac{\hbar^2}{m}\bigr)^2 \frac{1}{g} \langle \hat{O}_c \rangle$ by the equation of motion for thermodynamic states, the bilinear operator $\hat{O}_t$ will contribute to the full Wilson coefficients. Taken together, they contribute a finite term
\begin{align}
&W_c^{(3)}(\omega,{\bf q}) \nonumber \\
&\quad= W_t(\omega,{\bf q}) \biggl(\frac{\hbar^2}{m}\biggr)^2 \biggl[\frac{1}{g} - \frac{1}{{\cal A}^2} \times \raisebox{-0.35cm}{\scalebox{0.4}{\includegraphics{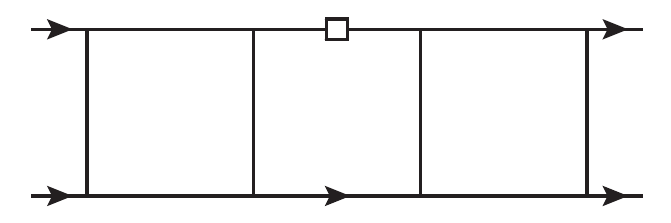}}}\biggr] \nonumber \\
&\quad\stackrel{\hbar\omega\to\varepsilon_{\bf q}}{\to} \frac{\hbar^2}{4\pi m a} \frac{1}{(\hbar \omega - \varepsilon_{\bf q})^2} .
\end{align}
The sum of the three contributions gives Eqs.~\eqref{eq:limitOPEBose} and~\eqref{eq:Pionshell}. For fermions, all symmetry factors work out in such a way that all coefficients are multiplied by a factor of $2$. For the full OPE beyond the single-particle peak, we refer the reader to Refs.~\cite{son10,hofmann11,goldberger12, nishida12}. Note that while for the IA, final-state corrections of order $\mathcal{O}(1/qa)$ are hard to compute, they are readily included in the OPE, essentially because the scattering between bosons with large momentum is the same as for free particles.

\section{Field theory of Bose-Einstein condensates}\label{app:bogoliubov}

Here, we discuss the field theory of a Bose-Einstein condensate and define the many-body $T$ matrix. The partition function of the Bose gas in terms of the coherent state path integral is given by
\begin{align}
\mathcal{Z} &= \int \mathcal{D}[\phi,\phi^*] \, e^{- S/\hbar} , \label{eq:partition}
\end{align}
where the imaginary time action reads
\begin{align}
S &= \int_0^{\hbar\beta} d\tau \int d{\bf r} \, \biggl\{ \phi^* \Bigl(\hbar \partial_\tau - \frac{\hbar^2 \nabla^2}{2m} - \mu\Bigr) \phi + \frac{g}{2} |\phi|^4\biggr\} . \label{eq:action}
\end{align}
The field $\phi$ has length dimension $L^{-3/2}$ and the coupling $g$ has dimension $\hbar^2 L/m$. As usual, we define the Fourier transform as
\begin{align}
f(\tau, {\bf x}) &= \frac{1}{\beta V} \sum_{i\omega_n, \bf k} e^{- i \omega_n \tau + i {\bf k} \cdot {\bf x}} \, f(i \omega_n, {\bf k}), \nonumber \\
f(i \omega_n, {\bf k}) &= \frac{1}{\hbar} \int_0^{\hbar\beta} d\tau \int d{\bf x} \, e^{i \omega_n \tau - i {\bf k} \cdot {\bf x}} \, f(\tau, {\bf x}) ,
\end{align}
except for the field $\phi({\bf x})$, which is expanded in a plane-wave basis as
\begin{align}
\phi(\tau, {\bf x}) &= \frac{1}{\hbar \beta \sqrt{V}} \sum_{i\omega_n, \bf k} e^{- i \omega_n \tau + i {\bf k} \cdot {\bf x}} \phi(i\omega_n, {\bf k}) , \nonumber \\
\phi(i\omega_n, {\bf k})  &= \frac{1}{\sqrt{V}} \int_0^{\hbar\beta} d\tau \int d{\bf x} \, e^{i \omega_n \tau - i {\bf k} \cdot {\bf x}} \phi({\bf x}) .
\end{align}
In the Bogoliubov approximation, we separate the condensate mode as
\begin{align}
\phi(\tau, {\bf x}) &= \phi_0 + \varphi(\tau, {\bf x}) \\
\phi^*(\tau, {\bf x}) &= \phi_0 + \varphi^*(\tau, {\bf x}) ,
\end{align}
where $\phi_0$ solves the Gross-Pitaevskii equation, which for a homogeneous condensate implies $\phi_0=\sqrt{n_0}$ and $\mu = gn_0$ (we set the phase of $\phi_0$ to zero). The action~[Eq.~\eqref{eq:action}] can then be written as a quadratic term $S_2$ in $\varphi$ and an interaction term $S_{\rm int}$, which collects higher powers of $\varphi$:
\begin{align}
S &= S_2 + S_{\rm int} ,
\end{align}
with
\begin{align}
S_2 &= \frac{1}{2 \hbar \beta} \sum_{i\omega_n, \bf k}
\begin{pmatrix} \varphi^*(i\omega_n, {\bf k}) &\varphi(-i\omega_n, -{\bf k}) \end{pmatrix} \nonumber \\
&\quad \times \begin{pmatrix}
-i\hbar\omega_n + \varepsilon_{\bf k} + g n_0 &g n_0 \\
g n_0 &i\hbar\omega_n + \varepsilon_{\bf k} + g n_0
\end{pmatrix} \nonumber \\
&\quad \times \begin{pmatrix} \varphi(i\omega_n, {\bf k}) \\ \varphi^*(-i\omega_n, -{\bf k}) \end{pmatrix}  \label{eq:S2}
\end{align}
and
\begin{align}
S_{\rm int} &= \int_0^{\hbar\beta} d\tau \int d{\bf r} \, \biggl\{ g \sqrt{n_0} \bigl(\varphi^{*2} \varphi + \varphi^{*} \varphi^2\bigr) + \frac{g}{2} \varphi^{*2} \varphi^2 \biggr\} .
\end{align}
The Feynman rules in momentum and frequency space are as follows: the single-particle propagator $G(i\omega_n,{\bf k})$ is read off directly from Eq.~\eqref{eq:S2} and is stated in Eq.~\eqref{eq:propagator} of the main text. In detail, we have
\begin{align}
G_{11}(i\omega_n,{\bf q}) &= \frac{u_{\bf q}^2}{i \hbar\omega_n - E_{\bf q}} - \frac{v_{\bf q}^2}{i \hbar\omega_n + E_{\bf q}} \\
G_{12}(i\omega_n,{\bf q}) &= - u_{\bf q} v_{\bf q} \biggl[\frac{1}{i \hbar\omega_n - E_{\bf q}} - \frac{1}{i \hbar\omega_n + E_{\bf q}}\biggr] \\
G_{22}(i\omega_n,{\bf q}) &= G_{11}^*(i\omega_n,{\bf q}) \\
G_{21}(i\omega_n,{\bf q}) &= G_{12}(i\omega_n,{\bf q}) 
\end{align}
with the Bogoliubov coherence factor of Eq.~\eqref{eq:coherencefactor}. The density insertion has unit matrix vertex $\sigma_0$ in Bogoliubov space, and the current insertion has matrix element $\frac{\hbar}{2 m} (2 {\bf k} + {\bf q}) \sigma_3$, where $\hbar {\bf q}$ is the momentum inserted by the current. We impose energy and momentum conservation at each vertex and integrate over every undetermined loop momentum with measure $\frac{1}{\beta V} \sum_{i\omega_n, {\bf k}}$.

%++++++++++++++++++++++++++++++++++++++++
\begin{figure}[t!]
\scalebox{0.6}{\includegraphics{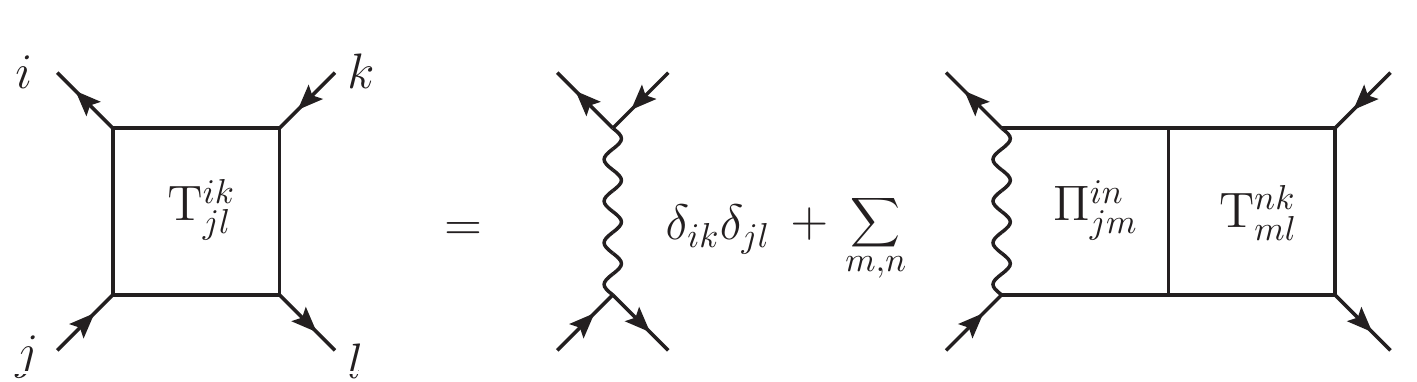}}
\caption{Bethe-Salpeter equation for the two-phonon $T$ matrix, indicated by a box. The wavy line denotes the contact interaction with strength $g$ and continuous lines indicate Bogoliubov propagators given in Eq.~\eqref{eq:propagator}. Indices are Bogoliubov indices.}
\label{fig:bethe}
\end{figure}
%++++++++++++++++++++++++++++++++++++++++

We now construct the many-body $T$ matrix within Bogoliubov theory. The Bogoliubov $T$ matrix has recently been used in many-body theories that predict the transition temperature and the condensate fraction of a weakly interacting BEC~\cite{watabe13}, and to formulate theories of the longitudinal susceptibility~\cite{watabe14} consistent with exact results for the infrared behavior~\cite{nepomnyashchii78,zwerger04,pistolesi04}. The $T$ matrix is a $4\times 4$ matrix, and we denote the components by $T_{jl}^{ik}(i\omega_n, {\bf q})$, where $k$ and $j$ are the ingoing Bogoliubov indices and $i$ and $l$ the outgoing indices, and $i\omega_n$ and ${\bf q}$ are the frequency and momentum transported through the diagram. The full matrix structure is
\begin{align}
T(i\omega_n, {\bf q}) &=
\begin{pmatrix}
T_{11}^{11} &T_{12}^{11} &T_{11}^{12} &T_{12}^{12} \\[1ex]
T_{21}^{11} &T_{22}^{11} &T_{21}^{12} &T_{22}^{12} \\[1ex]
T_{11}^{21} &T_{12}^{21} &T_{11}^{22} &T_{12}^{22} \\[1ex]
T_{21}^{21} &T_{22}^{21} &T_{21}^{22} &T_{22}^{22}
\end{pmatrix} .
\end{align}
The $T$ matrix is a solution of the Bethe-Salpeter equation diagrammatically shown in Fig.~\ref{fig:bethe}:
\begin{align}
T &= g + g \Pi T , \label{eq:Tmatrix}
\end{align}
where $\Pi$ is the one-loop bubble with components
\begin{align}
\Pi_{jl}^{ik}(i\omega_n, {\bf q}) &= \frac{1}{\beta V} \sum_{i\Omega_n, {\bf k}} G_{ki}(i \Omega_n + i \omega_n, {\bf k} + {\bf q}) G_{jl}(i \Omega_n, {\bf k}) .
\end{align}
The symmetries of the Bogoliubov Green's functions reduce the number of independent components:
\begin{align}
\Pi &= 
\begin{pmatrix}
\Pi_{a}^{} &\Pi_{b}^{} &\Pi_{b}^{*} &\Pi_{c}^{} \\
\Pi_{b}^{} &\Pi_{d}^{} &\Pi_{c}^{} &\Pi_{b}^{} \\
\Pi_{b}^{*} &\Pi_{c}^{} &\Pi_{d}^{*} &\Pi_{b}^{*} \\
\Pi_{c}^{} &\Pi_{b}^{} &\Pi_{b}^{*} &\Pi_{a}^{}
\end{pmatrix} ,
\end{align}
where
\begin{align}
\Pi_{a}(i\omega_n, {\bf q}) &= \frac{1}{V} \sum_{\bf k} \biggl\{\bigl[u_1^2 u_2^2 R_1(i\omega_n) + v_1^2 v_2^2 R_1(-i\omega_n)\bigr] \nonumber \\
&\quad- \bigl[u_1^2 v_2^2 R_2(i\omega_n) + v_1^2 u_2^2 R_2(-i\omega_n)\bigr]\biggr\} \label{eq:Pi11} \\
\Pi_{b}(i\omega_n, {\bf q}) &= \frac{1}{V} \sum_{\bf k} \biggl\{- u_1 v_1 \bigl[v_2^2 R_1(i\omega_n) + u_2^2 R_1(-i\omega_n)\bigr] \nonumber \\
&\quad+ u_1 v_1 \bigl[u_2^2 R_2(i\omega_n) + v_2^2 R_2(-i\omega_n)\bigr]\biggr\} ,
\end{align}
where $u_1 = u_{\bf k+q}$, $u_2 = u_{\bf k}$, $v_1 = v_{\bf k+q}$, $v_2 = v_{\bf k}$ as in the main text, $R_1$ and $R_2$ are defined in Eqs.~\eqref{eq:R1} and~\eqref{eq:R2}, and $\Pi_c$ and $\Pi_d$ are defined in Eqs.~\eqref{eq:Pic} and~\eqref{eq:Pid} of the main text. We renormalize the interaction in the standard way as given in Eq.~\eqref{eq:ren}. In particular, this implies that the $22$ and the $33$ components of $T^{-1} = \frac{1}{g} - \Pi$ remain finite, as the divergence of $\frac{1}{g}$ cancels a divergence in $\Pi_{22}$. The $11$ and the $44$ components of $T^{-1}$, however, diverge. Computing the $T$ matrix, we note that only the components with two ingoing or outgoing lines are nonzero:
\begin{align}
\begin{pmatrix}
T_{22}^{11} &T_{21}^{12} \\[1ex]
T_{12}^{21} &T_{11}^{22}
\end{pmatrix} 
&=
\frac{1}{(\frac{m}{8 \pi \hbar^2 a} - \Pi_{d}^*) (\frac{m}{8 \pi \hbar^2 a} - \Pi_{d}^{}) - \Pi_{c}^2} \nonumber \\
&\quad\times \begin{pmatrix}
\frac{m}{8 \pi \hbar^2 a} - \Pi_{d}^* &\Pi_{c}^{} \\[1ex]
\Pi_{c}^{} &\frac{m}{8 \pi \hbar^2 a} - \Pi_{d}^{}
\end{pmatrix} .
\end{align}
In the noncondensed phase, where $v\to 1$ and $u\to 0$, the off-diagonal terms vanish and the diagonal elements reduce to the standard $T$ matrix of a thermal gas.

\bibliography{bib}

\end{document}